 \documentclass[aps, twocolumn, showpacs, amsmath]{revtex4-2}
\usepackage{dcolumn}
\usepackage{bm}
\usepackage{graphicx}
\usepackage{color}
\usepackage{hyperref}
\usepackage{amsthm}
\usepackage{amssymb}

\begin{document}
\def\be{\begin{equation}}
\def\ee{\end{equation}}

\def\bc{\begin{center}}
\def\ec{\end{center}}
\def\bea{\begin{eqnarray}}
\def\eea{\end{eqnarray}}
\newcommand{\avg}[1]{\langle{#1}\rangle}
\newcommand{\Avg}[1]{\left\langle{#1}\right\rangle}

\def\ie{\textit{i.e.}}
\def\etal{\textit{et al.}}
\def\m{\vec{m}}
\def\G{\mathcal{G}}

\newcommand{\davide}[1]{{\bf\color{blue}#1}}
\newcommand{\gin}[1]{{\bf\color{green}#1}}

\title{Finite-time stochastic resetting in bounded domains: competing targets, energetic cost, and optimal search strategies}
\author{Hanshuang Chen}\email{chenhshf@ahu.edu.cn}
\author{Chongyao Tang}
\author{Guofeng Li}
\affiliation{School of Physics, Anhui University, Hefei 230601, China}
\begin{abstract}
We investigate the first-passage dynamics and energetic cost of a Brownian particle subject to finite-time stochastic resetting in a bounded one-dimensional domain with two absorbing boundaries. Unlike conventional instantaneous resetting, the particle is returned to a prescribed resetting position through a finite-duration process driven by a linear trapping potential. By solving the corresponding backward equations, we derive analytical expressions for the mean global first-passage time and the splitting probabilities. Our results show that the finite duration of the return process
can substantially modify the dependence of the mean global first-passage
time on the resetting rate. In particular, we identify a critical
trapping strength above which resetting can accelerate the search
process, and characterize how the range of favorable resetting
positions depends on the trapping strength. We further characterize the energetic cost of resetting by deriving the generating function of the total work and obtaining an analytical expression for its mean value. For the linear trapping potential considered here, we establish an exact relation between the mean total work and the mean cumulative return time. This relation allows us to establish a quantitative trade-off between the mean global first-passage time and the energetic expenditure, revealing how the finite duration of the return process affects search efficiency. Brownian dynamics simulations confirm our theoretical predictions.
\end{abstract}

\maketitle

\section{Introduction}

First-passage processes play a fundamental role in stochastic
dynamics and provide a natural framework for describing the time
required for a fluctuating system to reach a prescribed target for
the first time \cite{redner2001guide}. They arise in a wide range of physical, chemical,
biological, and financial processes, including diffusion-limited
reactions \cite{RevModPhys.85.135}, molecular transport \cite{bressloff2014stochastic}, neuronal firing \cite{PhysRevLett.93.048102}, animal foraging \cite{RevModPhys.83.81},
and the crossing of financial thresholds \cite{Valenti2007Hitting}. Understanding the mechanisms that control first-passage times
is therefore of considerable interest in nonequilibrium
statistical physics \cite{bray2013persistence}.

One particularly effective strategy for controlling first-passage
dynamics is stochastic resetting \cite{evans2020stochastic,rotbart2015michaelis,de2020optimization,752c-wqly}, whereby a diffusing particle
is returned to a prescribed position at random times. Since its
introduction into diffusion problems, resetting has attracted
considerable attention because it can dramatically alter the
first-passage properties of stochastic processes \cite{evans2011diffusion,evans2011diffusion2,pal2017first,chechkin2018random,PhysRevLett.122.020602}. In particular,
resetting can suppress long excursions away from a target and
render the mean first-passage time finite, or even reduce it
relative to unrestricted diffusion. The resulting search dynamics depend sensitively on the resetting rate, the initial
position, and the spatial configuration of the target. These
features have motivated extensive studies of resetting in
different geometries \cite{evans2014diffusion,Christou2015,Durang2019,PhysRevE.105.034109,3bc4-nlkt} and for a variety of stochastic processes \cite{PhysRevLett.112.240601,PhysRevLett.113.220602,evans2018run,bressloff2020occupation,kumar2020active,sytg-m4ms,PhysRevE.104.024105,PhysRevE.105.L012106,singh2020random,PhysRevLett.128.200603}, as well as various applications \cite{gupta2014fluctuating,meylahn2015large,PhysRevLett.130.157101,PhysRevE.103.052119,guo2023extremal,blumer2022stochastic,blumer2024combining}.

Most existing studies consider instantaneous resetting, in which
the particle is returned to the resetting position immediately
after a resetting event. Although this idealization provides a
mathematically convenient description, it neglects the finite
time and physical resources required to implement a resetting
protocol. In realistic systems, returning a particle to a
prescribed position generally requires an external force or
control mechanism, and the return process takes a nonzero
amount of time \cite{tal2020experimental,besga2020optimal,besga2021dynamical,goerlich2025taming}. 
More recently, finite-time resetting protocols have attracted increasing attention, in which a particle is returned to a prescribed position through an external trapping potential over a finite time interval \cite{evans2019effects,pal2019invariants,pal2019time,maso2019transport,bodrova2020resetting,bodrova2020brownian,pal2020search,gupta2020stochastic,gupta2021resetting,PhysRevE.110.L042101}. However, implementing such a return process generally requires an external control force and therefore comes at an energetic cost \cite{fuchs2016stochastic,sunil2023cost,olsen2024thermodynamic,goerlich2026time}. In stochastic thermodynamics, work is associated with changes in externally controlled parameters of a system \cite{seifert2012stochastic}. In particular, switching a trapping potential on and off generally involves work, and the energetic cost depends on the particle positions at the switching events \cite{gupta2020work,gupta2022work}. The interplay between the temporal and energetic costs of resetting therefore raises a natural question: how can one balance search efficiency against energy consumption in finite-time resetting processes? Indeed, the relation between the mean first-passage time and the energetic cost of resetting has recently been investigated in a one-dimensional semi-infinite geometry \cite{pal2023thermodynamic}, and a joint renewal formula for work and first-passage time in diffusion with stochastic return has been established, revealing the thermodynamic cost–time trade-off by optimizing the shape of the confining potential \cite{singh2024emerging}. However, the role of spatial confinement and competing absorbing targets in finite-time resetting remains less explored \cite{PhysRevE.102.022115}. In a bounded interval with two absorbing boundaries, the resetting position plays a crucial role in determining the search efficiency: resetting can either accelerate or hinder the search, depending on where the particle is reset to \cite{PhysRevE.99.032123}. This gives rise to the so-called ``resetting transition" characterized by a boundary separating the regions where resetting is beneficial from those where it is detrimental \cite{PhysRevLett.116.170601,pal2017first,PhysRevLett.113.220602,Christou2015,PhysRevE.97.062106,PhysRevResearch.1.032001,PhysRevE.105.034109}. An important question is how the finite duration of the return process modifies this transition and, in particular, how the boundary between these two regions depends on the strength of the trapping potential.

In this work, we extend the study of finite-time stochastic resetting to a bounded one-dimensional interval with two competing absorbing boundaries. Unlike the semi-infinite geometry considered in previous studies, the presence of two absorbing targets introduces a nontrivial dependence of the search efficiency on the resetting position. In particular, depending on the resetting position, resetting can either reduce or increase the mean first-passage time relative to ordinary diffusion, giving rise to a resetting transition between favorable and unfavorable regions. We investigate how the finite duration of the return process modifies this transition. By deriving the mean global first-passage time, we identify the conditions under which resetting accelerates the search and determine how the boundary separating the favorable and unfavorable regions depends on the trapping strength. We also calculate the splitting probability for reaching either absorbing boundary and show that it coincides with that of the corresponding instantaneous-resetting process, since absorption is inactive during the return phase.

We further investigate the energetic cost associated with finite-time resetting. Using a backward-equation approach, we derive the generating function of the work distribution and obtain the mean total work accumulated during the search process. For the linear trapping potential considered here, we establish an exact relation between the mean work and the mean return time, which is independent of the diffusivity during the return phase. Combining this result with the expression for the global first-passage time, we obtain an explicit time-work relation that quantifies the energetic cost of search acceleration. Our results thus reveal how finite-time resetting affects the resetting transition in a bounded domain and provide a thermodynamic characterization of the resulting search strategy. Finally, Brownian dynamics simulations are performed to verify our analytical predictions.

\section{Model setup}
We consider a Brownian particle undergoing a stochastic search process with finite-time resetting. The particle initially starts at position $x_0$ and performs free diffusion with diffusion coefficient $D$, which we refer to as the \textit{exploration} phase. The duration of each exploration phase is exponentially distributed, with probability density
\begin{eqnarray}\label{eq1.0}
f_R(t)=re^{-rt},
\end{eqnarray}
where $r$ is the resetting rate. 

\begin{figure}
	\centerline{\includegraphics*[width=1.0\columnwidth]{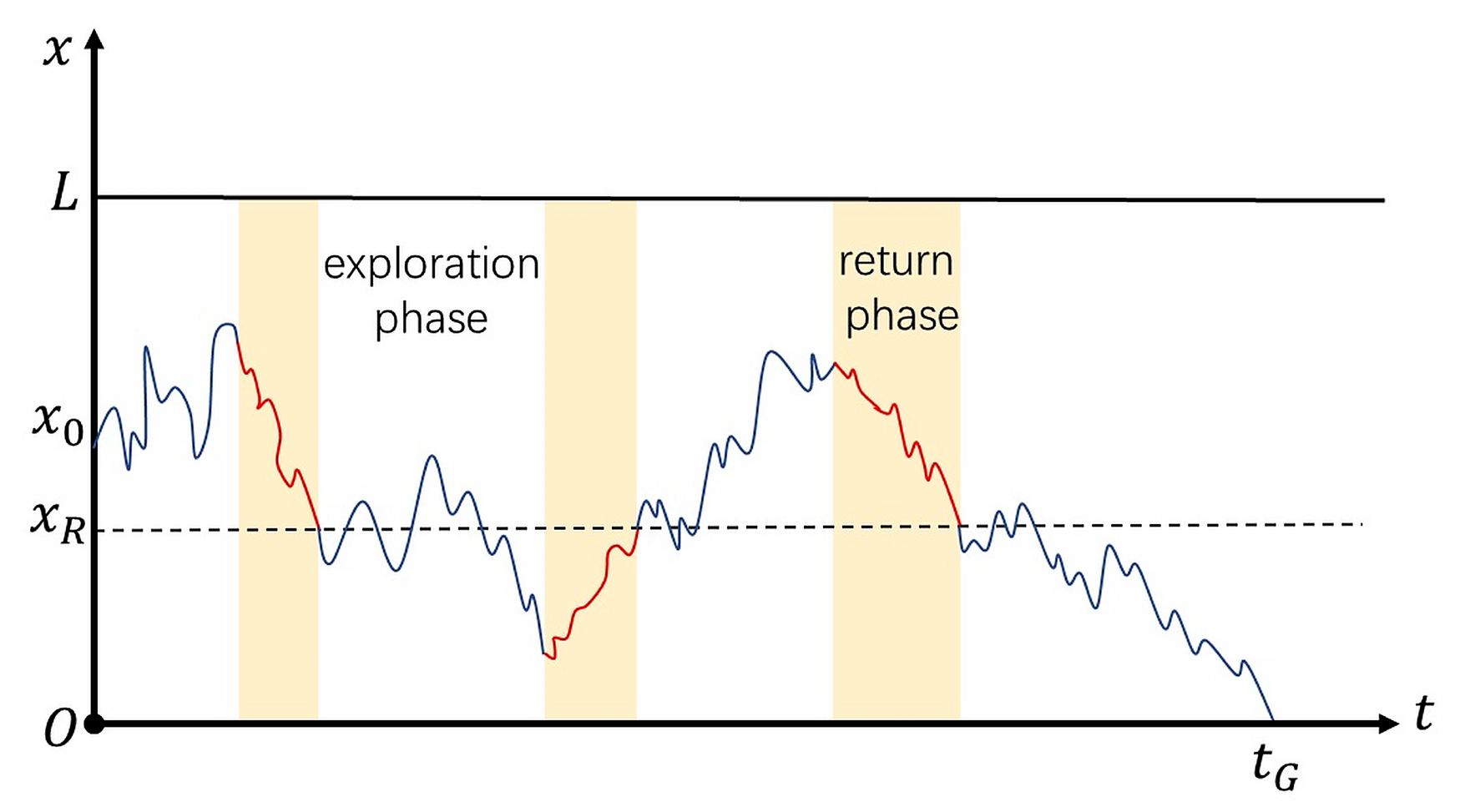}}
	\caption{Schematic illustration of finite-time resetting in a one-dimensional Brownian search process. The particle initially undergoes free Brownian motion during the exploration phase (blue trajectories), starting from $x_0$ in the interval $0<x<L$. At a prescribed resetting time, an external potential is switched on to drive the particle toward the reset position $x_R$ (red trajectories).
		Once the particle reaches $x_R$, the external potential is switched off, and a new exploration phase begins. The yellow shaded regions indicate the finite-duration resetting phases. Both boundaries, $x=0$ and $x=L$, are perfectly absorbing, and the process terminates when the particle first reaches either boundary. \label{fig0}}
\end{figure}

During the exploration phase, the particle is absorbed immediately upon reaching either of the two boundaries, $x=0$ or $x=L$. If a resetting event occurs before absorption, an external trapping potential centered at a prescribed resetting position $x_R\in(0,L)$ is switched on, driving the particle back toward $x_R$. During this \textit{return} phase, the diffusion coefficient changes from $D$ to $D_R$. Importantly, the absorbing boundaries are inactive during the return phase. Thus, the particle is allowed to cross either boundary without being absorbed and continues its motion under the trapping potential until it first reaches $x_R$. Upon reaching $x_R$, the trapping potential is switched off, and the particle resumes free diffusion in the interval $\left( 0,L\right) $. The exploration-return cycle is repeated until the particle reaches either absorbing boundary during an exploration phase (see Fig.~\ref{fig0} for an illustration).

The dynamics can be described by the Langevin equation \cite{pal2023thermodynamic}
\begin{eqnarray}\label{eq1.1}
\dot{x} = -{\partial_x}U( x; h(t) )+\sqrt {2 D(t) } {\eta }(t),
\end{eqnarray}
where $\eta(t)$ is Gaussian white noise satisfying $\langle \eta(t) \rangle=0$ and $\langle\eta(t)\eta(t')\rangle=\delta(t-t')$. The time-dependent potential is given by
\begin{eqnarray}\label{eq1.1.1}
U( x; h(t) )=h(t)U(x),
\end{eqnarray}
where $h(t)$ is a dichotomous variable that takes the value 0 during exploration and 1 during the return phase. Accordingly, the diffusion coefficient switches between $D$ and $D_R$ in the two phases, respectively. We assume a linear trapping potential,
\begin{eqnarray}\label{eq1.2}
U(x)=a |x-x_R|,
\end{eqnarray}
where $a>0$ characterizes the trap strength.

The distinction between the two phases is essential to the definition of the search process: absorption can occur only during exploration, whereas the return dynamics takes place on the entire real line in the presence of the trapping potential. In particular, the return phase is terminated exclusively by the first arrival at $x_R$, irrespective of whether the particle crosses either of the two target boundaries during its return.

The main objective of this work is twofold. First, we investigate the global first-passage time $t_G$ required for the particle to reach either target and determine the conditions under which this time can be optimized by tuning the resetting rate $r$. The global first-passage time can be decomposed into the cumulative times spent in the exploration and return phases:
\begin{eqnarray}\label{eq1.3}
t_G=t_E+t_R,
\end{eqnarray}
where $t_E$ and $t_R$ denote the total exploration time and the total return time, respectively. Since the exploration phases are interrupted by exponentially distributed resetting events, the cumulative exploration time is equivalent to the first-passage time of the corresponding instantaneous-resetting process, provided that both processes have the same resetting protocol and initial condition. In contrast, $t_R$ represents the additional time cost associated with finite-time resetting.

Second, we investigate the thermodynamic cost associated with the resetting process, characterized by the work performed through switching the trapping potential on and off. In the instantaneous-resetting limit, corresponding to an infinitely strong trapping potential, the work required to return the particle to $x_R$ generally diverges. For finite-time resetting, however, the trapping strength remains finite, resulting in a finite mean work at the expense of an additional return time. These competing effects give rise to a trade-off between the search time and the thermodynamic cost, which constitutes another central focus of this work.

\section{Mean first-passage time}

Let $Q_A(p|x_0)$ denote the Laplace transform of the probability
density function (PDF) of the global first-passage time $t_G$,
defined as
\begin{eqnarray}\label{eq2.1}
	Q_A(p|x_0)
	=\int_0^\infty dt_G\,e^{-pt_G}P_A(t_G|x_0)
	\equiv \left\langle e^{-pt_G}\right\rangle,
\end{eqnarray}
where $P_A(t_G|x_0)$ is the PDF of $t_G$. Once $Q_A(p|x_0)$
is determined, the moments of the global first-passage time
can be obtained through
\begin{eqnarray}\label{eq2.1.1}
	\left\langle t_G^n (x_0)\right\rangle
	=(-1)^n\left.
	\frac{\partial^n Q_A(p|x_0)}{\partial p^n}
	\right|_{p=0}.
\end{eqnarray}

Following the Brownian functional approach \cite{majumdar2005brownian,singh2022first,pal2023thermodynamic}, the backward equation governing $Q_A(p|x_0)$ can
be written as
\begin{eqnarray}\label{eq2.2}
	\left[D\partial_{x_0}^2-(p+r)\right]Q_A(p|x_0)+rQ_R(p|x_0)Q_A(p|x_R)=0, \nonumber \\
\end{eqnarray}
where $Q_R(p|x_0)$ is the Laplace transform of the first-passage-time
PDF associated with the return phase, defined by
\begin{eqnarray}\label{eq2.3}
	Q_R(p|x_0)
	=\int_0^\infty dt_R\,e^{-pt_R}P_R(t_R|x_0),
\end{eqnarray}
with $P_R(t_R|x_0)$ denoting the PDF of the return time $t_R$.

For the linear trapping potential specified in Eq.~(\ref{eq1.2}),
the return-time Laplace transform can be obtained analytically
(see, e.g., Ref.~\cite{pal2023thermodynamic}), yielding
\begin{eqnarray}\label{eq2.4}
	Q_R(p|x_0)
	=e^{-\lambda(p)|x_0-x_R|},
\end{eqnarray}
where
\begin{eqnarray}\label{eq2.5}
	\lambda(p)
	=\frac{\sqrt{a^2+4pD_R}-a}{2D_R}.
\end{eqnarray}
Interestingly, the mean return time,
\begin{eqnarray}
	\left\langle t_R(x_0)\right\rangle
	=-\left.\partial_p Q_R(p|x_0)\right|_{p=0}
	=\frac{|x_R-x_0|}{a},
\end{eqnarray}
is independent of the diffusion coefficient $D_R$. In the limit
of an infinitely strong trapping potential, $a\to\infty$, the
mean return time vanishes, recovering the instantaneous-resetting
limit.

Substituting Eq.~(\ref{eq2.4}) into Eq.~(\ref{eq2.2}), we obtain
a differential equation for $Q_A(p|x_0)$. Since the return-time
transform has different functional forms on the two sides of
$x_R$, we solve the equation separately for $x_0>x_R$ and
$x_0<x_R$. The resulting general solution is
\begin{widetext}
\begin{eqnarray}\label{eq2.6}
Q_A(p|x_0)
=
\begin{cases}
C_1e^{\mu(p)x_0}+C_2e^{-\mu(p)x_0}
-\nu(p)Q_A(p|x_R)
e^{-\lambda(p)(x_0-x_R)},
&x_0>x_R,
\\[2mm]
C_3e^{\mu(p)x_0}+C_4e^{-\mu(p)x_0}
-\nu(p)Q_A(p|x_R)
e^{-\lambda(p)(x_R-x_0)},
&x_0<x_R,
\end{cases}
\end{eqnarray}
\end{widetext}
where
\begin{eqnarray}
	\mu(p)=\sqrt{\frac{p+r}{D}},
	\qquad
	\nu(p)=\frac{r}{D\lambda^2(p)-(p+r)}.
\end{eqnarray}
The four integration constants $C_1$, $C_2$, $C_3$, and $C_4$
are determined by the absorbing boundary conditions and the
matching conditions at $x_0=x_R$:
\begin{eqnarray}
	Q_A(p|0)&=&Q_A(p|L)=1,
	\\
	Q_A(p|x_R^+)&=&Q_A(p|x_R^-),
	\\
	\left.\partial_{x_0}Q_A(p|x_0)\right|_{x_0\to x_R^+}
	&=&
	\left.\partial_{x_0}Q_A(p|x_0)\right|_{x_0\to x_R^-}.
\end{eqnarray}
Here, the boundary conditions $Q_A(p|0)=Q_A(p|L)=1$ follow
from the fact that the global first-passage time is zero when
the particle starts at either absorbing boundary. The matching
conditions ensure the continuity of the solution and its first
spatial derivative at the resetting position.

Setting $x_0=x_R$ in Eq.~(\ref{eq2.6}), we obtain
\begin{eqnarray}
	Q_A(p|x_R)
	=
	\frac{C_{1,3}e^{\mu(p)x_R}
		+C_{2,4}e^{-\mu(p)x_R}}
	{1+\nu(p)},
\end{eqnarray}
where $C_{1,3}$ and $C_{2,4}$ denote the corresponding integration
constants on the two sides of $x_R$. After applying the boundary
and matching conditions, we obtain the following expression for
the special case in which the initial position coincides with
the resetting position:
\begin{eqnarray}\label{eq2.7}
	Q_A(p|x_R)
	=
	\frac{
		\mu e^{\lambda L+(\lambda+\mu)x_R}
		(e^{\mu L}-1)(e^{\mu L}+e^{2\mu x_R})
	}{
		f_A(p;x_R)
	},
\end{eqnarray}
where the dependence of $\lambda$, $\mu$, and $\nu$ on $p$
is implicit, and
\begin{eqnarray}\label{eq2.8}
	f_A(p;x_R)
	&=&
	\lambda\mu e^{\lambda(L+x_R)+2\mu L}
	+\lambda\mu e^{\lambda(L+x_R)+4\mu x_R}
	\nonumber\\
	&&+\mu\nu e^{\mu(L+x_R)+2\lambda x_R}
	-\mu\nu e^{\lambda L+2\mu L+\mu x_R}
	\nonumber\\
	&&+\mu\nu e^{\lambda L+3\mu x_R}
	-\mu\nu e^{2\lambda x_R+\mu L+3\mu x_R}
	\nonumber\\
	&&+(\mu-\lambda\nu+\mu\nu)
	e^{(\lambda+2\mu)(L+x_R)}
	\nonumber\\
	&&-(\mu+\lambda\nu+\mu\nu)
	e^{\lambda(L+x_R)+2\mu x_R}.
\end{eqnarray}

The mean global first-passage time is then obtained by differentiating
Eq.~(\ref{eq2.7}) with respect to $p$ and taking the limit $p\to0$:
\begin{eqnarray}\label{eq2.9}
	\left\langle t_G\right\rangle
	=-\left.\partial_p Q_A(p|x_R)\right|_{p=0}
	=\left\langle t_E\right\rangle
	+\left\langle t_R\right\rangle.
\end{eqnarray}
The mean global first-passage time consists of two contributions:
the cumulative exploration time and the cumulative return time.
In particular, the former is given by
\begin{eqnarray}\label{eq2.10}
	\left\langle t_E\right\rangle
	=
	\frac{
		(e^{\alpha x_R}-1)(e^{\alpha L}-e^{\alpha x_R})
	}{
		r(e^{\alpha L}+e^{2\alpha x_R})
	},
\end{eqnarray}
where
\begin{eqnarray}
	\alpha=\sqrt{\frac{r}{D}}.
\end{eqnarray}
This expression coincides with the mean first-passage time
for instantaneous resetting to $x_R$ \cite{PhysRevE.99.032123}. The additional contribution
arising from finite-time resetting is
\begin{eqnarray}\label{eq2.11.0}
\langle t_R\rangle=\frac{\Delta_G(\alpha;x_R)}{a},
\end{eqnarray}
where
	\begin{eqnarray}\label{eq2.11}
		\Delta_G(\alpha;x_R)
		&=&
		\frac{
			(e^{2\alpha x_R}-e^{2\alpha L})
			\left(1-e^{2\alpha x_R}
			+\alpha x_Re^{\alpha x_R}\right)
		}{
			\alpha e^{\alpha x_R}
			(e^{\alpha L}-1)(e^{\alpha L}+e^{2\alpha x_R})
		}
		\nonumber\\
		&&+
		\frac{
			\alpha(L-x_R)e^{\alpha(L+x_R)}
			(1-e^{2\alpha x_R})
		}{
			\alpha e^{\alpha x_R}
			(e^{\alpha L}-1)(e^{\alpha L}+e^{2\alpha x_R})
		}.
	\end{eqnarray}
Clearly, $\langle t_R\rangle>0$ for any finite trapping strength
$a>0$, whereas this contribution vanishes in the limit $a\to\infty$.

\begin{figure}
	\centerline{\includegraphics*[width=1.0\columnwidth]{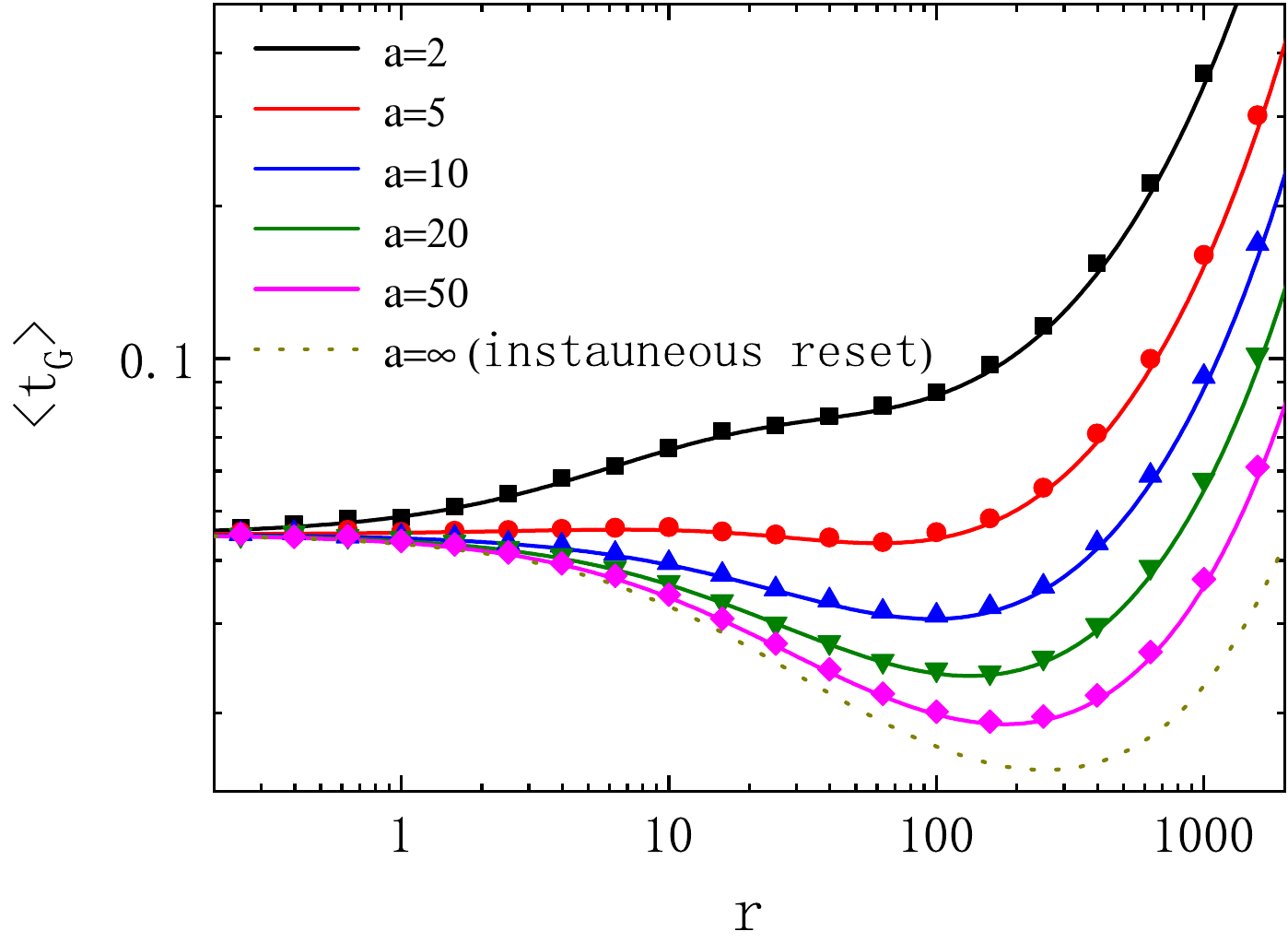}}
	\caption{Mean global first-passage time $\langle t_G\rangle$ as a
		function of the resetting rate $r$ for different trapping strengths
		$a$. The limit $a\to\infty$ corresponds to instantaneous resetting.
		Solid lines represent the theoretical predictions, while symbols
		denote the Brownian dynamics simulation results. The other parameters
		are $x_0=x_R=0.1$, $L=1$, and $D=D_R=1$. \label{fig1}}
\end{figure}

Figure~\ref{fig1} illustrates the dependence of the mean global
first-passage time $\langle t_G\rangle$ on the resetting
rate $r$ for different trapping strengths $a$, with other parameters fixed: $x_0=x_R=0.1$, $L=1$, and $D=D_R=1$. For
sufficiently strong trapping, the mean first-passage time
exhibits a nonmonotonic dependence on $r$: it initially
decreases with increasing resetting rate, reaches a
minimum at an optimal resetting rate, $r=r_{{\rm{opt}}}$, and subsequently
increases. This behavior indicates that resetting can
accelerate the search when the return process is sufficiently
fast. As the trapping strength increases, the optimal
resetting rate shifts and the results approach the
instantaneous-resetting limit. By contrast, for weak trapping, the mean global
first-passage time increases monotonically with $r$,
indicating that the additional time spent in the return
phases outweighs the benefit of restarting the exploration
process.  To validate our theoretical results, we conduct numerical simulations to compute the mean global first-passage time. In the simulations, the Langevin dynamics was discretized using a time step of $dt=10^{-5}$, and all the data were obtained by averaging over $10^5$ independent realizations. The simulation data (denoted by symbols in Fig.~\ref{fig1}) are in excellent agreement with the theoretical predictions.

We next investigate how the resetting position affects the mean global first-passage time. To this end, we first recall the main results for instantaneous resetting \cite{PhysRevE.99.032123},
corresponding to the limit $a\to\infty$. In this limit, the mean
first-passage time is determined entirely by the exploration
process. Depending on the resetting position, $\langle t_E\rangle$
exhibits either nonmonotonic or monotonically increasing behavior
as a function of the resetting rate $r$. Specifically, when the
dimensionless resetting position
\begin{eqnarray}
	\bar{x}_R=\frac{x_R}{L}
\end{eqnarray}
satisfies
\begin{eqnarray}
	\bar{x}_R\in
	\left(0,\frac{5-\sqrt{5}}{10}\right)
	\cup
	\left(\frac{5+\sqrt{5}}{10},1\right),
\end{eqnarray}
the mean exploration time is nonmonotonic in $r$ and attains
a minimum at an intermediate resetting rate. Thus, instantaneous
resetting can accelerate the search process when the resetting
position is sufficiently close to either absorbing boundary.
By contrast, when $\bar{x}_R\in
\left(\frac{5-\sqrt{5}}{10}, \frac{5+\sqrt{5}}{10}\right)$ 
the mean exploration time increases monotonically with $r$,
indicating that resetting is detrimental to the search process.

For finite-time resetting, the additional return-time contribution
$\Delta_G/a$ increases the global first-passage time relative to
the instantaneous-resetting case. Consequently, the parameter
region in which resetting can reduce the mean search time is
expected to shrink. This raises two natural questions: under what
conditions can finite-time resetting still accelerate the search
process, and how strong must the trapping potential be for such
an acceleration to occur?

To address these questions, we examine the dependence of $\langle t_G\rangle$ on the resetting rate $r$. In particular,
we evaluate its derivative in the weak-resetting limit $r\to0$.
A negative value of $\frac{\partial\langle t_G\rangle}{\partial r}
	|_{r\to0}$ indicates that introducing a small resetting rate initially
reduces the mean global first-passage time. This analysis therefore
provides a necessary condition for resetting to be beneficial.
The resulting critical resetting positions are
\begin{eqnarray}\label{eq2.12}
	\bar{x}_R^\pm
	=
	\frac{1}{2}
	\pm
	\frac{
		\sqrt{5\bar{a}^2+32\bar{a}-64}
	}{
		2(5\bar{a}-8)
	},
\end{eqnarray}
where
\begin{eqnarray}
	\bar{a}=\frac{aL}{D}
\end{eqnarray}
is the dimensionless trapping strength. In the limit
$\bar{a}\to\infty$, these critical positions approach $ 	\bar{x}_R^\pm\to\frac{5\pm\sqrt{5}}{10}$,
recovering the instantaneous-resetting result. As $\bar{a}$
decreases, $\bar{x}_R^+$ shifts toward larger values, whereas
$\bar{x}_R^-$ shifts toward smaller values. This indicates that
the range of resetting positions for which resetting can
accelerate the search becomes progressively narrower as the
trapping strength decreases.

The two critical positions cease to define an interval within
the physical domain $0\leq\bar{x}_R\leq1$ when the dimensionless
trapping strength falls below
\begin{eqnarray}\label{eq2.13}
	\bar{a}_c=4.
\end{eqnarray}
Thus, the weak-resetting criterion predicts that a finite
resetting rate can initially reduce the mean global first-passage
time only when $\bar{a}>4$ and the resetting position belongs to
\begin{eqnarray}
	\bar{x}_R\in(0,\bar{x}_R^-)
	\cup(\bar{x}_R^+,1).
\end{eqnarray}
In contrast, for $\bar{a}\leq4$, no resetting position within
the interval satisfies this criterion.

\begin{figure}
	\centerline{\includegraphics*[width=1.0\columnwidth]{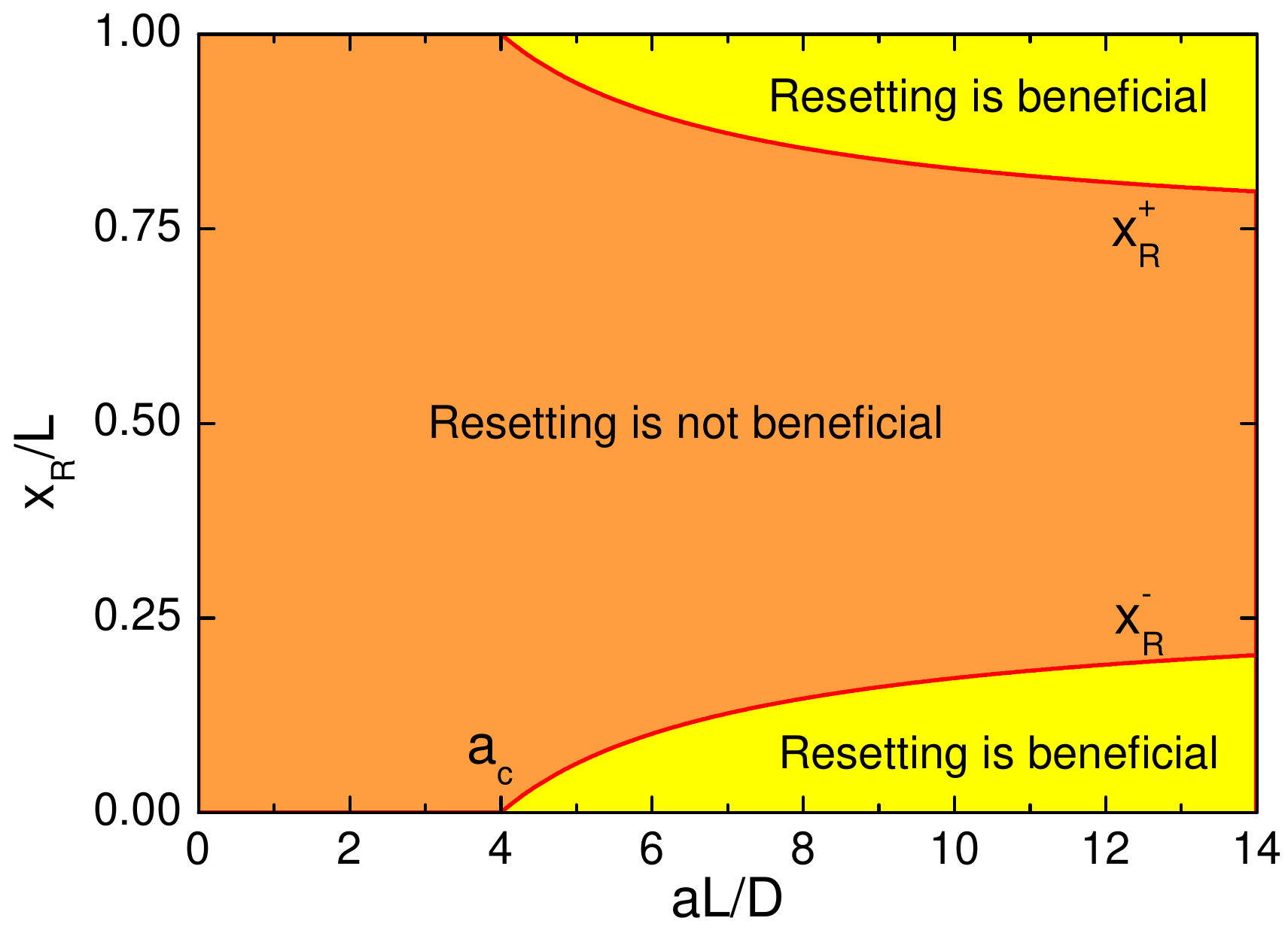}}
	\caption{Phase diagram showing the parameter regions in which
		finite-time resetting is beneficial, as determined by the
		weak-resetting criterion. The solid curves indicate the critical
		resetting positions $ \bar{x}_R^- $ and $ \bar{x}_R^+ $, which
		separate the regions where resetting reduces or increases the mean
		global first-passage time. The critical dimensionless trapping
		strength is $\bar{a}_c=4$, with $\bar{a}=aL/D$ and
		$\bar{x}_R=x_R/L$. \label{fig2}}
\end{figure}

Figure~\ref{fig2} summarizes the parameter regions in which
finite-time resetting is beneficial. It shows
that the favorable region is located near the two
absorbing boundaries, whereas resetting is detrimental
when the resetting position lies in the central region
of the interval. As the dimensionless trapping strength
$\bar a$ increases, the favorable regions expand toward
the interior of the interval. In particular, the critical
trapping strength is $\bar a_c=4$, below which no
resetting position satisfies the weak-resetting
acceleration criterion. In the limit $\bar a\to\infty$,
the phase boundaries approach those of instantaneous
resetting.

\section{Splitting Probability}

In addition to the global first-passage time, it is also
important to characterize the probability that the particle
eventually reaches one target before the other. Specifically, we denote by $\Pi_R(x_0)$ the splitting probability that a particle starting from $x_0$ is eventually absorbed at the right boundary, while the corresponding probability for the left boundary is given by $\Pi_L(x_0)=1-\Pi_R(x_0)$.

During an exploration phase, the particle undergoes free diffusion with diffusion coefficient $D$, while resetting events occur at rate $r$. If a resetting event occurs before absorption, the particle enters the return phase and is driven toward the resetting position $x_R$. Both absorbing boundaries are inactive during this phase, and exploration resumes from $x_R$ once the return is completed. As a consequence, the splitting probability is identical to that of instantaneous resetting to the same position, being independent of the duration and detailed dynamics of the return process. We therefore present its derivation below for completeness.

By considering the evolution over an infinitesimal time
interval, the backward equation for $\Pi_R(x_0)$ reads \cite{guo2023extremal}
\begin{equation}
D\frac{\mathrm{d}^2\Pi_R(x_0)}{\mathrm{d}x_0^2}
-r\Pi_R(x_0)
+r\Pi_R(x_R)=0,
\quad 0<x_0<L.
\label{eq:splitting_backward}
\end{equation}
The boundary conditions are
\begin{equation}
\Pi_R(0)=0,
\qquad
\Pi_R(L)=1.
\label{eq:splitting_bc}
\end{equation}
Here, the term $r\Pi_R(x_R)$ accounts for the fact that,
following a resetting event, the return process is completed
without absorption and the particle resumes exploration
from $x_R$.

\begin{figure}
	\centerline{\includegraphics*[width=1.0\columnwidth]{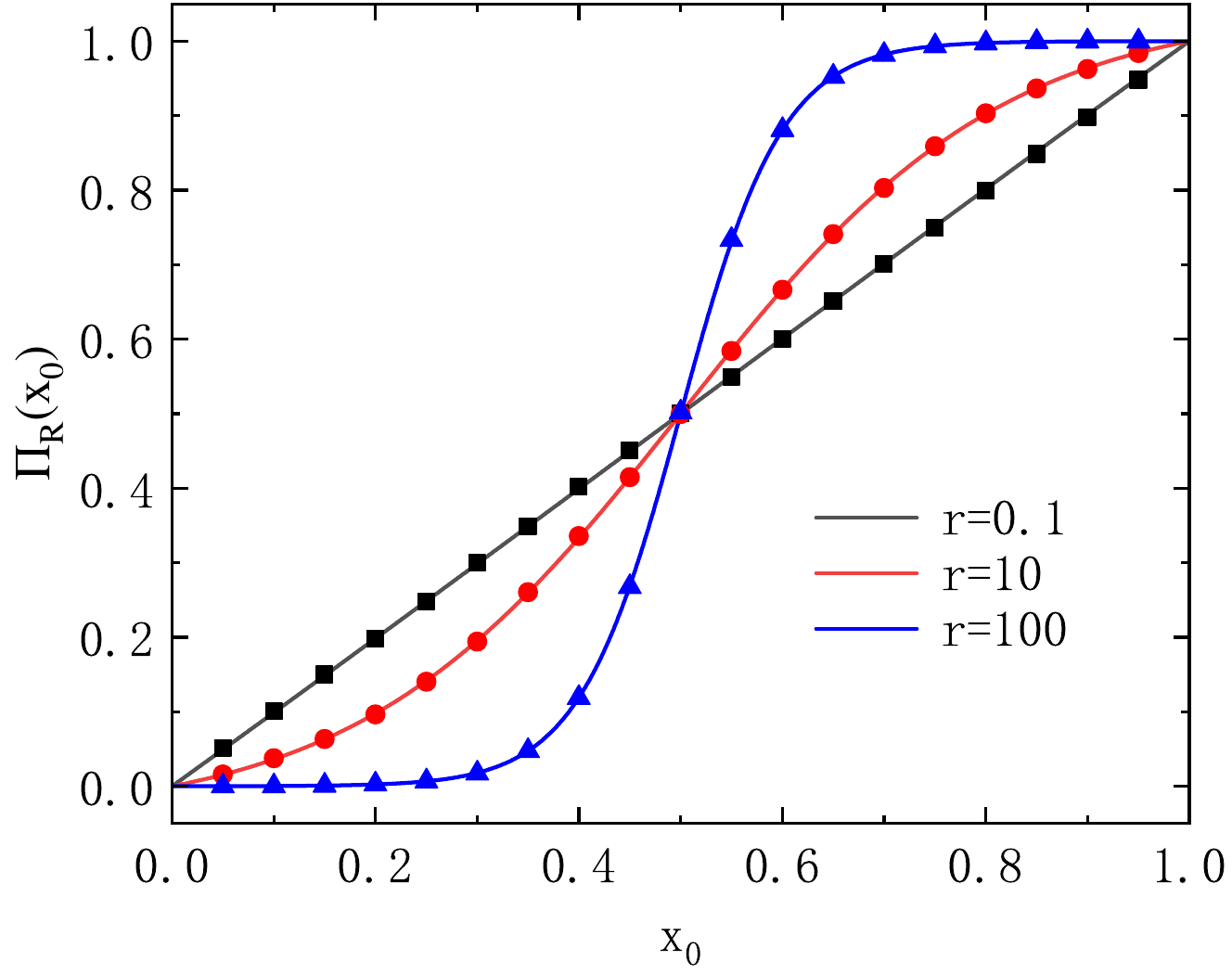}}
	\caption{Splitting probability $\Pi_R(x_0)$ as a
		function of $x_0$ for different resetting rates
		$r$. Solid lines represent the theoretical predictions, while symbols
		denote the Brownian dynamics simulation results. The other parameters
		are $x_R=x_0$, $L=1$, and $D=D_R=1$. \label{figsp}}
\end{figure}

The solution of Eq.~(\ref{eq:splitting_backward}) can be
written as
\begin{equation}
\Pi_R(x_0)
=
\Pi_R(x_R)
\left[
1-\cosh(\alpha x_0)
+\coth(\alpha x_R)\sinh(\alpha x_0)
\right].
\end{equation}
where again $\alpha=\sqrt{r/D}$. Using the boundary condition at $x=L$, we obtain
\begin{equation}
\Pi_R(x_R)
=
\frac{1}{
	1-\cosh(\alpha L)+\coth(\alpha x_R)\sinh(\alpha L)
}.
\end{equation}
Consequently, the splitting probability is given explicitly by
\begin{equation}
	\Pi_R(x_0)
	=
	\frac{
		1-\cosh(\alpha x_0)
		+\coth(\alpha x_R)\sinh(\alpha x_0)
	}{
		1-\cosh(\alpha L)
		+\coth(\alpha x_R)\sinh(\alpha L)
	}.
\label{eq:splitting_probability}
\end{equation}
The splitting probability for the left boundary follows
from $\Pi_L(x_0)=1-\Pi_R(x_0)$.

Figure~\ref{figsp} illustrates the splitting probability $\Pi_R(x_0)$ versus the starting position $x_0$ under different resetting rates $r$. When the resetting rate takes the limit $r\to 0$, the curve is linear, $\lim_{r\to 0}\Pi_R(x_0)={x_0}/{L}$, recovering the splitting probability of ordinary Brownian motion. As the resetting rate increases, the curve gradually develops a sharp, step-like shape: for large $r$, the splitting probability stays close to zero for small $x_0$ and rapidly rises to unity once $x_0$ exceeds $L/2$. The solid lines are analytical theoretical results, and discrete symbols are Brownian dynamics simulations, showing good agreement.

\section{Mean total work for resetting}

We next investigate the thermodynamic cost associated with the
resetting process. Within the framework of stochastic
thermodynamics, the work performed on the system due to the
time-dependent modulation of the external potential
$U(x;h(t))$ is defined as \cite{sekimoto2010stochastic}
\begin{eqnarray}\label{eq3.1}
	W=\int_0^{t_G}dt\,
	\frac{\partial U(x;h)}{\partial h}\dot{h}.
\end{eqnarray}
For the potential considered here,
$U(x;h)=hU(x)$, we have
$\partial_h U(x;h)=U(x)$. Moreover, since the dichotomous
variable $h(t)$ switches between $0$ and $1$, its time
derivative can be expressed as
\begin{eqnarray}
	\dot{h}(t)
	=
	\sum_i\delta(t-t_i)
	-
	\sum_j\delta(t-\tau_j),
\end{eqnarray}
where $t_i$ denotes the time at which the trapping potential
is switched on during the $i$th resetting event, while
$\tau_j$ denotes the time at which it is switched off after
the particle returns to $x_R$. Substituting this expression
into Eq.~(\ref{eq3.1}), we obtain 
\begin{eqnarray}\label{eq3.2}
	W
	&=&
	\sum_i U(x(t_i))
	-
	\sum_j U(x(\tau_j))
	\nonumber\\
	&=&
	\sum_i U(x(t_i))\nonumber\\ 
	&=& a \sum_i |x_i-x_R| .
\end{eqnarray}
In the second equality, we have used the fact that the return
phase terminates when the particle reaches $x_R$, where
$U(x_R)=0$. Thus, switching off the trapping potential
does not contribute to the work. Consequently, the total
work is entirely determined by the particle positions at
the instants when the trapping potential is switched on \cite{gupta2022work}.

To characterize the statistics of the total work, we
introduce its Laplace transform
\begin{eqnarray}\label{eq3.3}
	Q_C(p|x_0)
	=
	\int_0^\infty dW\,e^{-pW}P_C(W|x_0)
	\equiv
	\left\langle e^{-pW}\right\rangle,
\end{eqnarray}
where $P_C(W|x_0)$ is the PDF
of the total work $W$ for a particle starting at $x_0$.
Following the Brownian functional approach, the backward
equation governing $Q_C(p|x_0)$ can be written as \cite{pal2023thermodynamic}
\begin{eqnarray}\label{eq3.4}
	\left[D\partial_{x_0}^2-r\right]Q_C(p|x_0)
	+r e^{-pU(x_0)}Q_C(p|x_R)=0.
\end{eqnarray}
The factor $e^{-pU(x_0)}$ accounts for the work performed
when the trapping potential is switched on at the position
$x_0$. Since the work accumulated during the subsequent
return phase is already accounted for by the resetting
cycle, no additional work contribution arises during the
return itself.

For the linear trapping potential given in Eq.~(\ref{eq1.2}),
we have $U(x_0)=a|x_0-x_R|$. Equation~(\ref{eq3.4}) can
therefore be solved separately in the regions $x_0>x_R$
and $x_0<x_R$. The general solution is
\begin{widetext}
	\begin{eqnarray}\label{eq3.5}
		Q_C(p|x_0)
		=
		\begin{cases}
			C_1e^{\alpha x_0}+C_2e^{-\alpha x_0}
			-\gamma(p)Q_C(p|x_R)e^{-pa(x_0-x_R)},
			& x_0>x_R,
			\\[2mm]
			C_3e^{\alpha x_0}+C_4e^{-\alpha x_0}
			-\gamma(p)Q_C(p|x_R)e^{-pa(x_R-x_0)},
			& x_0<x_R,
		\end{cases}
	\end{eqnarray}
	where
	\begin{eqnarray}
		\alpha=\sqrt{\frac{r}{D}},
		\qquad
		\gamma(p)=\frac{r}{D p^2a^2-r}.
	\end{eqnarray}
\end{widetext}
The four integration constants are determined by the
absorbing boundary conditions and the matching conditions
at $x_0=x_R$:
\begin{eqnarray}
	Q_C(p|0)&=&Q_C(p|L)=1,
	\\
	Q_C(p|x_R^+)&=&Q_C(p|x_R^-),
	\\
	\left.\partial_{x_0}Q_C(p|x_0)\right|_{x_0\to x_R^+}
	&=&
	\left.\partial_{x_0}Q_C(p|x_0)\right|_{x_0\to x_R^-}.
\end{eqnarray}
The first two conditions follow from the fact that no
additional work is performed once the particle reaches
either absorbing boundary. The remaining conditions ensure
the continuity of the solution and its first spatial
derivative at the resetting position. Setting $x_0=x_R$
in Eq.~(\ref{eq3.5}), we can determine $Q_C(p|x_R)$
self-consistently. In particular, for the case in which
the initial position coincides with the resetting position,
we obtain
\begin{eqnarray}\label{eq3.6}
	Q_C(p|x_R)
	=
	\frac{
		e^{ap(L-x_R)+\alpha x_R}
		(e^{\alpha L}-1)
		(e^{\alpha L}+e^{2\alpha x_R})
	}{
		f_C(p;x_R)
	},
\end{eqnarray}
where
\begin{eqnarray}\label{eq3.7}
	f_C(p;x_R)
	&=&
	a\alpha^{-1}p\gamma(p)
	\left[
	e^{ap(L-x_R)+2\alpha L}
	+
	e^{ap(L-x_R)+4\alpha x_R}
	\right]
	\nonumber\\
	&&+\gamma(p)
	\left[
	e^{\alpha(L+x_R)}
	-e^{\alpha(L+3x_R)}
	\right.
	\nonumber\\
	&&\left.
	+e^{ap(L-2x_R)+3\alpha x_R}
	-e^{ap(L-2x_R)+\alpha(2L+x_R)}
	\right]
	\nonumber\\
	&&+
	\left[
	1+\gamma(p)-a\alpha^{-1}p\gamma(p)
	\right]
	\nonumber\\
	&&\times
	\left[
	e^{ap(L-x_R)+2\alpha(L+x_R)}
	-e^{ap(L-x_R)+2\alpha x_R}
	\right].
\end{eqnarray}

The mean total work can be obtained from the first derivative
of $Q_C(p|x_R)$ with respect to $p$ at $p=0$:
\begin{eqnarray}\label{eq3.8.0}
	\left\langle W\right\rangle
	=
	-\left.\partial_p Q_C(p|x_R)\right|_{p=0}
	=
	a \Delta_G(\alpha;x_R),
\end{eqnarray}
where $\Delta_G(\alpha;x_R)$ is given in Eq.~\ref{eq2.11}. 

Equation~(\ref{eq3.8.0}) shows that the mean total work
is proportional to the trapping strength $a$, whereas
the function $\Delta_G(\alpha;x_R)$ depends on the resetting
rate through $\alpha=\sqrt{r/D}$ and on the resetting
position $x_R$. In particular, the mean work diverges
in the limit $a\to\infty$, corresponding to instantaneous
resetting. This divergence reflects the increasingly
large work required to return the particle to $x_R$
over an increasingly short time.

In the limit of a semi-infinite domain, $L\to\infty$,
Eq.~(\ref{eq3.8.0}) reduces to
\begin{eqnarray}
	\left\langle W\right\rangle_{L\to\infty}
	=
	\frac{a}{\alpha}
	\left[
	2\sinh(\alpha x_R)-\alpha x_R
	\right],
\end{eqnarray}
which recovers the corresponding result for resetting
in a semi-infinite interval \cite{pal2023thermodynamic}.

\begin{figure}
	\centerline{\includegraphics*[width=1.0\columnwidth]{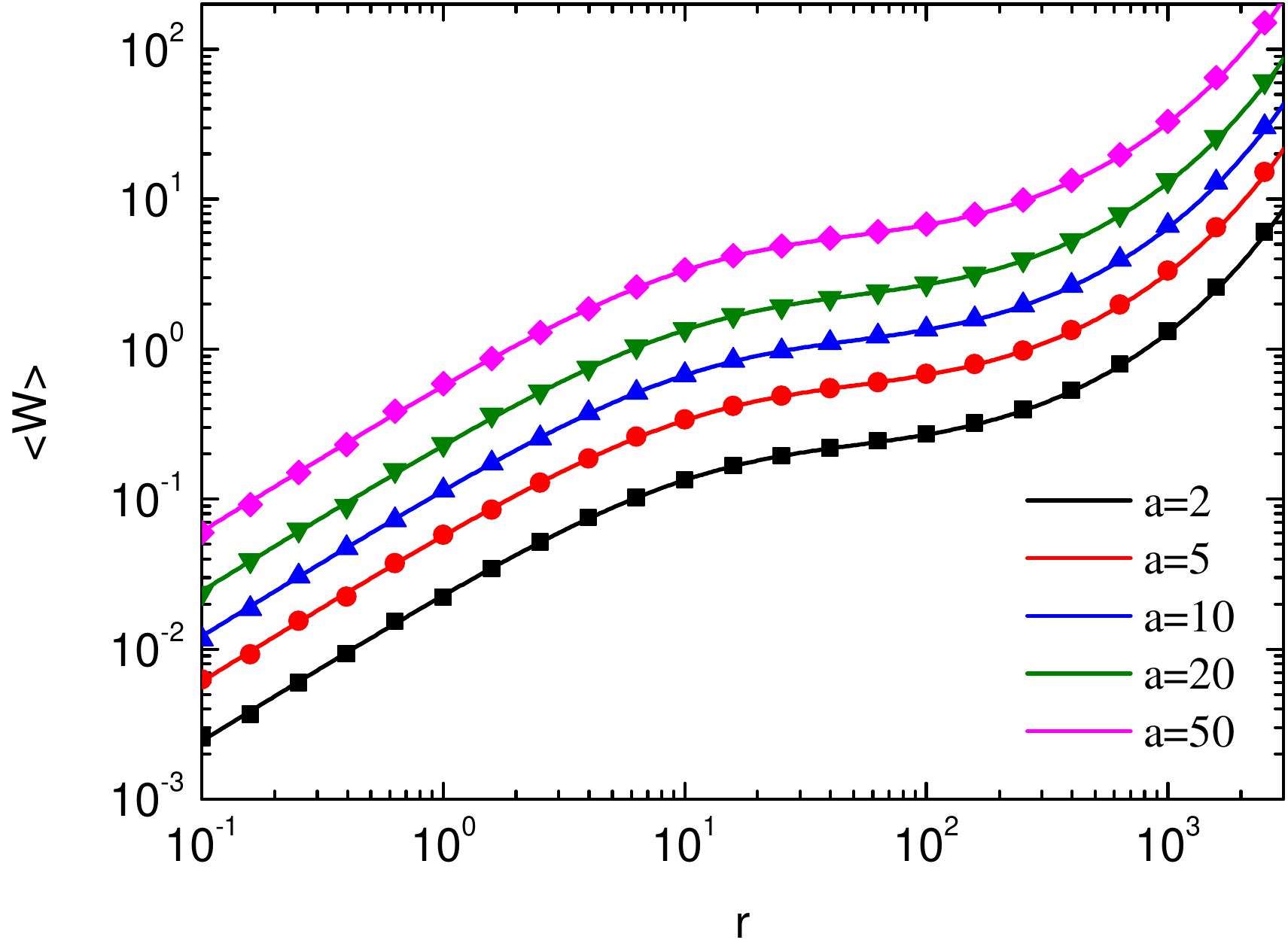}}
	\caption{Mean total work $\langle W\rangle$ as a function of the
		resetting rate $r$ for different trapping strengths $a$. Both axes
		are plotted on logarithmic scales. Solid lines represent the
		theoretical predictions, while symbols denote the Brownian dynamics
		simulation results. The other parameters are $x_R=0.1$, $L=1$,
		and $D=D_R=1$.  \label{fig3}}
\end{figure}

Figure~\ref{fig3} shows the dependence of the mean total work $\langle W\rangle$ on the resetting rate $r$ for different trapping strengths $a$. It is evident that the mean work increases monotonically with the resetting rate for all the considered values of $a$. This behavior can be understood from the fact that a larger resetting rate leads to more frequent resetting events, thereby increasing the cumulative energetic expenditure associated with switching on the trapping potential. Moreover, at a fixed resetting rate, the mean work increases with the trapping strength, as a stronger potential requires more work to initiate each return process. The theoretical predictions (solid lines) are in excellent agreement with the Brownian dynamics simulation results (symbols), confirming the validity of our analytical expression. These results demonstrate that increasing either the resetting frequency or the trapping strength enhances the energetic cost of the resetting protocol.

More importantly, the above results reveal an exact relation between the mean total work and the mean cumulative return time. Using Eqs.~(\ref{eq2.11.0}) and (\ref{eq3.8.0}), we obtain
\begin{eqnarray}
\langle W\rangle=a^2\langle t_R\rangle.
\end{eqnarray}
It is worth emphasizing that this exact relation originates from the specific linear form of the trapping potential, $U(x)=a|x-x_R|$, and does not generally hold for arbitrary trapping potentials. In fact, for a particle located at $x$ when the potential is switched on, the work required to initiate the return phase is $W(x)=a|x-x_R|$, which is proportional to both the trapping strength $a$ and the distance from the resetting position $x_R$. Meanwhile, the corresponding mean return time is $\langle t_R(x)\rangle=|x-x_R|/a$, which is proportional to the same distance but inversely proportional to $a$. The common dependence on $|x-x_R|$ therefore leads directly to the exact proportionality between the mean work and the mean cumulative return time, $\langle W\rangle=a^2\langle t_R\rangle$, after averaging over all resetting events. This relation holds for arbitrary resetting rates and resetting positions, and is independent of the return-phase diffusion coefficient $D_R$, under the present return dynamics.

\section{Trade-off relation}

The results obtained above allow us to establish a quantitative
relation between the mean global first-passage time and the
thermodynamic cost of resetting. Combining Eqs.~\ref{eq2.9} and
\ref{eq3.8.0}, and eliminating the trap strength $a$, we obtain
\begin{eqnarray}
	\label{eq4.1}
	\langle t_G \rangle
	=
	\langle t_E \rangle
	+
	\frac{\Delta_G^2 (\alpha;x_R)}{\langle W\rangle}.
\end{eqnarray}
This relation explicitly connects the mean search time to the
mean work required for finite-time resetting. The first term,
$\langle t_E\rangle$, is the cumulative exploration time and
coincides with the mean first-passage time of the corresponding
instantaneous-resetting process. The second term represents the
additional time cost associated with the finite duration of the
return phase. In particular, for fixed resetting rate and resetting
position, this contribution is inversely proportional to the mean
work. Thus, reducing the thermodynamic cost by weakening the
trapping potential inevitably increases the time spent in the
return phase.

Equation \ref{eq4.1} also reveals a trade-off between the mean
search time and the thermodynamic cost: decreasing the mean work
increases the return-time contribution, whereas increasing the
trapping strength reduces this contribution at the expense of
a larger work input. This relation highlights the competition
between search efficiency and energetic cost in finite-time
resetting.

To quantify the excess time induced by finite-time resetting, we
define
\begin{eqnarray}
	\label{eq4.2}
	\langle t_{\rm ex}\rangle
	\equiv
	\langle t_G\rangle
	-
	\left.\langle t_E\rangle\right|_{r=r_E^*},
\end{eqnarray}
where $r_E^*$ denotes the optimal resetting rate that minimizes
the mean first-passage time for instantaneous resetting. Substituting
Eq.~\ref{eq4.1} into this definition gives
\begin{eqnarray}
	\label{eq4.3}
	\langle t_{\rm ex}\rangle
	=
	\left[
	\langle t_E\rangle
	-
	\left.\langle t_E\rangle\right|_{r=r_E^*}
	\right]
	+
	\frac{\Delta_G ^2(\alpha;x_R)}{\langle W\rangle}.
\end{eqnarray}
The two contributions have distinct physical origins. The first
term measures the excess exploration time arising from a resetting
rate that differs from the optimal rate for instantaneous resetting.
The second term accounts for the additional time required to return
the particle to the resetting position. Since $r_E^*$ minimizes
$\langle t_E\rangle$, both contributions are nonnegative. Therefore,
$\langle t_{\rm ext}\rangle$ provides a natural measure of the
time penalty associated with finite-time resetting, relative to
the optimal instantaneous-resetting protocol.

\begin{figure}
	\centerline{\includegraphics*[width=1.0\columnwidth]{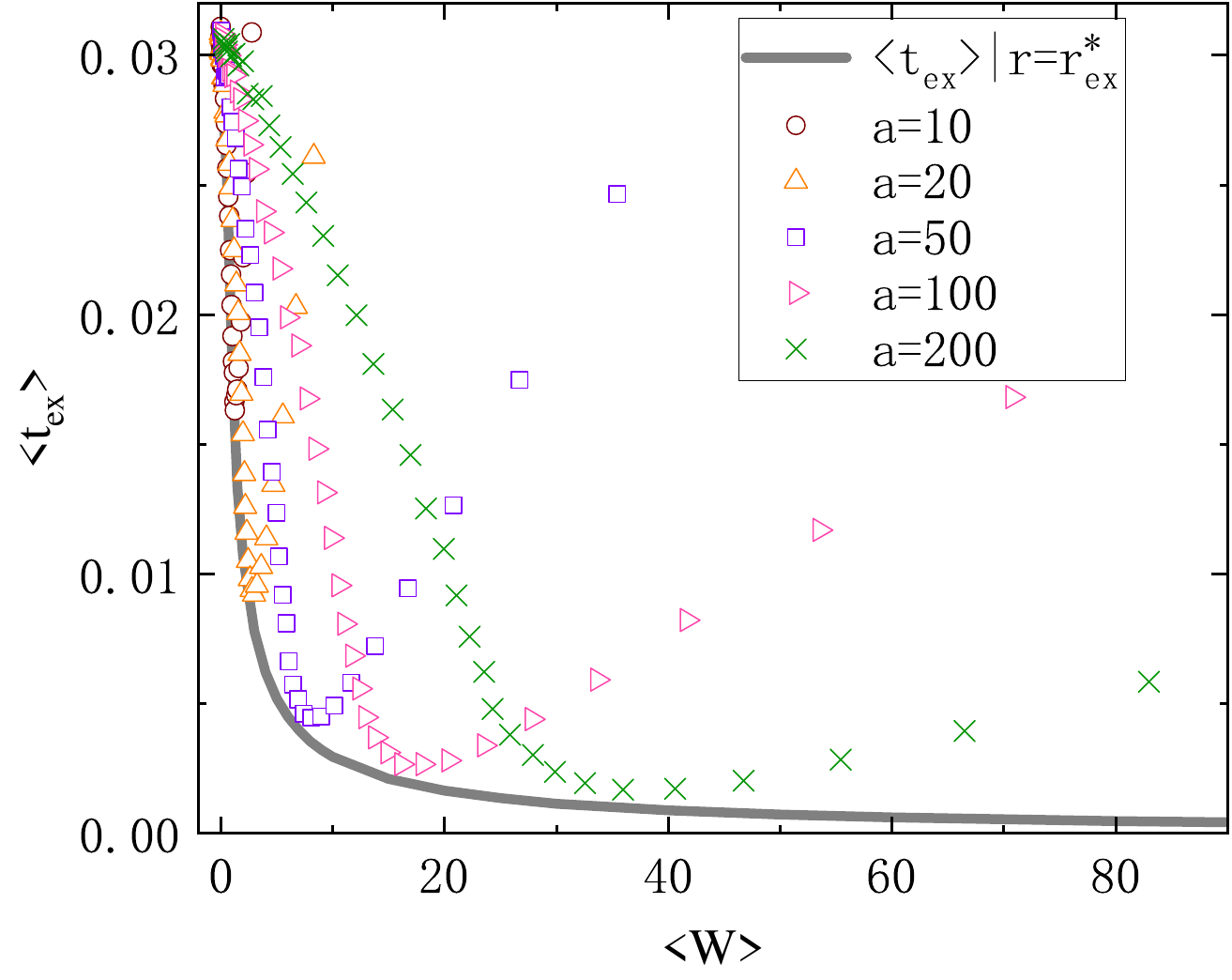}}
	\caption{Trade-off relation between excess time and work, i.e., mean excess global first-passage time	$\langle t_{\mathrm{ex}}\rangle$ as a function of the mean work
		$\langle W\rangle$ for different trapping strengths $a$. Solid
		lines represent the	minimum excess time $\langle t_{\text{ex}} \rangle\big|_{r=r_{\text{ex}}^*}$ for a given mean work, while symbols denote
		the Brownian dynamics simulation results. The other parameters are
		$x_R=0.1$, $L=1$, and $D=D_R=1$.  \label{fig4}}
\end{figure}

We solve $\partial_r \langle t_{\text{ex}} \rangle = 0$ to get the resetting rate $r = r_{\text{ex}}^*$ that minimizes $\langle t_{\text{ex}} \rangle$. The minimal excess time $\langle t_{\text{ex}} \rangle\big|_{r=r_{\text{ex}}^*}$ is shown by the black solid line versus $\langle W \rangle$ in Fig.~\ref{fig4}, which serves as the lower bound of $\langle t_{\text{ex}} \rangle$. To verify this, we performed numerical simulations
to illustrate the trade-off between
the excess global first-passage time and the mean
energetic cost. Each point there for a given resetting rate and $a$ is denoted by a symbol. In particular, the reference curve obtained
at $r=r_{\text{ex}}^*$ provides a lower bound for the excess
time at a given mean work, since the exploration-time
contribution is minimized at the optimal resetting
rate of the instantaneous-resetting process.

\section{CONCLUSION}

In this work, we have investigated the first-passage dynamics and energetic cost of a Brownian particle undergoing finite-time stochastic resetting in a bounded one-dimensional domain. In contrast to conventional instantaneous resetting, the protocol considered here returns the particle to a prescribed resetting position through a finite-duration process driven by a linear trapping potential. This introduces an additional temporal cost and a nonzero work input, providing a framework for examining the interplay between search efficiency and energetic expenditure.

We first derived the Laplace transform of the global first-passage-time distribution and obtained an analytical expression for the mean global first-passage time. The latter consists of two physically distinct contributions: the cumulative exploration time and the cumulative return time. The exploration-time contribution is identical to the MFPT of the corresponding instantaneous-resetting process, whereas the return-time contribution accounts for the finite duration of the resetting process. Our results demonstrate that finite-time resetting does not necessarily improve search efficiency. Whether resetting reduces the mean global first-passage time depends on both the trapping strength and the resetting position. In particular, a sufficiently strong trapping potential is required for a favorable weak-resetting regime to exist, and the range of resetting positions for which resetting is beneficial expands as the trapping strength increases. In the limit of infinitely strong trapping, the return duration vanishes and the instantaneous-resetting behavior is recovered.

We also derived the splitting probabilities for absorption at the two boundaries. An interesting feature of the present protocol is that these probabilities are independent of both the trapping strength and the diffusion coefficient during the return phase. This follows from the assumption that absorption is inactive during the return process: regardless of the duration or detailed trajectory of a return, the particle always resumes exploration from the same resetting position. Thus, while the return dynamics affect the temporal cost of the search, they do not alter the relative probabilities of reaching the two targets.

We next quantified the energetic cost associated with switching the trapping potential on and off. By deriving the generating function of the total work, we obtained an analytical expression for the mean total switching work. For the linear trapping potential considered here, the mean work increases linearly with the trapping strength, whereas the cumulative return time decreases inversely with it. These results lead to an exact relation, $\langle W\rangle=a^2\langle t_R\rangle$, which connects the mean work input directly to the mean cumulative return time. This relation originates from the fact that the work required to initiate an individual return is proportional to the distance from the resetting position, while the corresponding mean return time is proportional to the same distance but inversely proportional to the trapping strength. Importantly, the relation is independent of the return-phase diffusion coefficient under the present dynamics, although it is specific to the linear trapping potential and need not hold for general return protocols.

By eliminating the trapping strength, we further established a quantitative relation between the mean global first-passage time and the mean work. This relation makes explicit the competition between the temporal and energetic costs of finite-time resetting: increasing the work input by strengthening the trapping potential reduces the time spent in the return phases, whereas weakening the trap lowers the work cost but prolongs the return process. The resulting trade-off shows that optimizing the mean search time and minimizing the energetic expenditure are, in general, distinct objectives.

Overall, our findings provide an analytical characterization of finite-time resetting in a bounded domain with competing absorbing targets, highlighting how the return dynamics influence search times and energetic costs in different ways. The present framework can be extended in several directions, including nonlinear trapping potentials, higher-dimensional search geometries, and non-Brownian stochastic dynamics. Such extensions may help clarify how the geometry of the search domain and the physical mechanism of return jointly determine the efficiency and energetic requirements of stochastic resetting.

\begin{acknowledgments}
	This work is supported by the National Natural Science Foundation of China (Grant No. 11875069), the Key Scientific Research Fund of Anhui Provincial Education Department (Grant No. 2023AH050116), and Anhui Project (Grant No. 2022AH020009).
\end{acknowledgments}


\begin{thebibliography}{69}
	\expandafter\ifx\csname natexlab\endcsname\relax\def\natexlab#1{#1}\fi
	\expandafter\ifx\csname bibnamefont\endcsname\relax
	\def\bibnamefont#1{#1}\fi
	\expandafter\ifx\csname bibfnamefont\endcsname\relax
	\def\bibfnamefont#1{#1}\fi
	\expandafter\ifx\csname citenamefont\endcsname\relax
	\def\citenamefont#1{#1}\fi
	\expandafter\ifx\csname url\endcsname\relax
	\def\url#1{\texttt{#1}}\fi
	\expandafter\ifx\csname urlprefix\endcsname\relax\def\urlprefix{URL }\fi
	\providecommand{\bibinfo}[2]{#2}
	\providecommand{\eprint}[2][]{\url{#2}}
	
	\bibitem[{\citenamefont{Redner}(2001)}]{redner2001guide}
	\bibinfo{author}{\bibfnamefont{S.}~\bibnamefont{Redner}},
	\emph{\bibinfo{title}{A guide to first-passage processes}}
	(\bibinfo{publisher}{Cambridge University Press}, \bibinfo{year}{2001}).
	
	\bibitem[{\citenamefont{Bressloff and Newby}(2013)}]{RevModPhys.85.135}
	\bibinfo{author}{\bibfnamefont{P.~C.} \bibnamefont{Bressloff}}
	\bibnamefont{and} \bibinfo{author}{\bibfnamefont{J.~M.} \bibnamefont{Newby}},
	\bibinfo{journal}{Rev. Mod. Phys.} \textbf{\bibinfo{volume}{85}},
	\bibinfo{pages}{135} (\bibinfo{year}{2013}).
	
	\bibitem[{\citenamefont{Bressloff}(2014)}]{bressloff2014stochastic}
	\bibinfo{author}{\bibfnamefont{P.~C.} \bibnamefont{Bressloff}},
	\emph{\bibinfo{title}{Stochastic processes in cell biology}},
	vol.~\bibinfo{volume}{41} (\bibinfo{publisher}{Springer Berlin},
	\bibinfo{year}{2014}).
	
	\bibitem[{\citenamefont{Schindler et~al.}(2004)\citenamefont{Schindler,
			Talkner, and H{\"a}nggi}}]{PhysRevLett.93.048102}
	\bibinfo{author}{\bibfnamefont{M.}~\bibnamefont{Schindler}},
	\bibinfo{author}{\bibfnamefont{P.}~\bibnamefont{Talkner}}, \bibnamefont{and}
	\bibinfo{author}{\bibfnamefont{P.}~\bibnamefont{H{\"a}nggi}},
	\bibinfo{journal}{Phys. Rev. Lett.} \textbf{\bibinfo{volume}{93}},
	\bibinfo{pages}{048102} (\bibinfo{year}{2004}).
	
	\bibitem[{\citenamefont{B\'enichou et~al.}(2011)\citenamefont{B\'enichou,
			Loverdo, Moreau, and Voituriez}}]{RevModPhys.83.81}
	\bibinfo{author}{\bibfnamefont{O.}~\bibnamefont{B\'enichou}},
	\bibinfo{author}{\bibfnamefont{C.}~\bibnamefont{Loverdo}},
	\bibinfo{author}{\bibfnamefont{M.}~\bibnamefont{Moreau}}, \bibnamefont{and}
	\bibinfo{author}{\bibfnamefont{R.}~\bibnamefont{Voituriez}},
	\bibinfo{journal}{Rev. Mod. Phys.} \textbf{\bibinfo{volume}{83}},
	\bibinfo{pages}{81} (\bibinfo{year}{2011}).
	
	\bibitem[{\citenamefont{Valenti et~al.}(2007)\citenamefont{Valenti, Spagnolo,
			and Bonanno}}]{Valenti2007Hitting}
	\bibinfo{author}{\bibfnamefont{D.}~\bibnamefont{Valenti}},
	\bibinfo{author}{\bibfnamefont{B.}~\bibnamefont{Spagnolo}}, \bibnamefont{and}
	\bibinfo{author}{\bibfnamefont{G.}~\bibnamefont{Bonanno}},
	\bibinfo{journal}{Physica A} \textbf{\bibinfo{volume}{382}},
	\bibinfo{pages}{311} (\bibinfo{year}{2007}).
	
	\bibitem[{\citenamefont{Bray et~al.}(2013)\citenamefont{Bray, Majumdar, and
			Schehr}}]{bray2013persistence}
	\bibinfo{author}{\bibfnamefont{A.~J.} \bibnamefont{Bray}},
	\bibinfo{author}{\bibfnamefont{S.~N.} \bibnamefont{Majumdar}},
	\bibnamefont{and} \bibinfo{author}{\bibfnamefont{G.}~\bibnamefont{Schehr}},
	\bibinfo{journal}{Adv. Phys.} \textbf{\bibinfo{volume}{62}},
	\bibinfo{pages}{225} (\bibinfo{year}{2013}).
	
	\bibitem[{\citenamefont{Evans et~al.}(2020)\citenamefont{Evans, Majumdar, and
			Schehr}}]{evans2020stochastic}
	\bibinfo{author}{\bibfnamefont{M.~R.} \bibnamefont{Evans}},
	\bibinfo{author}{\bibfnamefont{S.~N.} \bibnamefont{Majumdar}},
	\bibnamefont{and} \bibinfo{author}{\bibfnamefont{G.}~\bibnamefont{Schehr}},
	\bibinfo{journal}{J. Phys. A: Math. Theor.} \textbf{\bibinfo{volume}{53}},
	\bibinfo{pages}{193001} (\bibinfo{year}{2020}).
	
	\bibitem[{\citenamefont{Rotbart et~al.}(2015)\citenamefont{Rotbart, Reuveni,
			and Urbakh}}]{rotbart2015michaelis}
	\bibinfo{author}{\bibfnamefont{T.}~\bibnamefont{Rotbart}},
	\bibinfo{author}{\bibfnamefont{S.}~\bibnamefont{Reuveni}}, \bibnamefont{and}
	\bibinfo{author}{\bibfnamefont{M.}~\bibnamefont{Urbakh}},
	\bibinfo{journal}{Phys. Rev. E} \textbf{\bibinfo{volume}{92}},
	\bibinfo{pages}{060101} (\bibinfo{year}{2015}).
	
	\bibitem[{\citenamefont{De~Bruyne et~al.}(2020)\citenamefont{De~Bruyne,
			Randon-Furling, and Redner}}]{de2020optimization}
	\bibinfo{author}{\bibfnamefont{B.}~\bibnamefont{De~Bruyne}},
	\bibinfo{author}{\bibfnamefont{J.}~\bibnamefont{Randon-Furling}},
	\bibnamefont{and} \bibinfo{author}{\bibfnamefont{S.}~\bibnamefont{Redner}},
	\bibinfo{journal}{Phys. Rev. Lett.} \textbf{\bibinfo{volume}{125}},
	\bibinfo{pages}{050602} (\bibinfo{year}{2020}).
	
	\bibitem[{\citenamefont{Biswas et~al.}(2025)\citenamefont{Biswas, Majumdar, and
			Pal}}]{752c-wqly}
	\bibinfo{author}{\bibfnamefont{A.}~\bibnamefont{Biswas}},
	\bibinfo{author}{\bibfnamefont{S.~N.} \bibnamefont{Majumdar}},
	\bibnamefont{and} \bibinfo{author}{\bibfnamefont{A.}~\bibnamefont{Pal}},
	\bibinfo{journal}{Phys. Rev. Lett.} \textbf{\bibinfo{volume}{135}},
	\bibinfo{pages}{227101} (\bibinfo{year}{2025}).
	
	\bibitem[{\citenamefont{Evans and
			Majumdar}(2011{\natexlab{a}})}]{evans2011diffusion}
	\bibinfo{author}{\bibfnamefont{M.~R.} \bibnamefont{Evans}} \bibnamefont{and}
	\bibinfo{author}{\bibfnamefont{S.~N.} \bibnamefont{Majumdar}},
	\bibinfo{journal}{Phys. Rev. Lett.} \textbf{\bibinfo{volume}{106}},
	\bibinfo{pages}{160601} (\bibinfo{year}{2011}{\natexlab{a}}).
	
	\bibitem[{\citenamefont{Evans and
			Majumdar}(2011{\natexlab{b}})}]{evans2011diffusion2}
	\bibinfo{author}{\bibfnamefont{M.~R.} \bibnamefont{Evans}} \bibnamefont{and}
	\bibinfo{author}{\bibfnamefont{S.~N.} \bibnamefont{Majumdar}},
	\bibinfo{journal}{J. Phys. A: Math. Theor.} \textbf{\bibinfo{volume}{44}},
	\bibinfo{pages}{435001} (\bibinfo{year}{2011}{\natexlab{b}}).
	
	\bibitem[{\citenamefont{Pal and Reuveni}(2017)}]{pal2017first}
	\bibinfo{author}{\bibfnamefont{A.}~\bibnamefont{Pal}} \bibnamefont{and}
	\bibinfo{author}{\bibfnamefont{S.}~\bibnamefont{Reuveni}},
	\bibinfo{journal}{Phys. Rev. Lett.} \textbf{\bibinfo{volume}{118}},
	\bibinfo{pages}{030603} (\bibinfo{year}{2017}).
	
	\bibitem[{\citenamefont{Chechkin and Sokolov}(2018)}]{chechkin2018random}
	\bibinfo{author}{\bibfnamefont{A.}~\bibnamefont{Chechkin}} \bibnamefont{and}
	\bibinfo{author}{\bibfnamefont{I.}~\bibnamefont{Sokolov}},
	\bibinfo{journal}{Phys. Rev. Lett.} \textbf{\bibinfo{volume}{121}},
	\bibinfo{pages}{050601} (\bibinfo{year}{2018}).
	
	\bibitem[{\citenamefont{Pal et~al.}(2019{\natexlab{a}})\citenamefont{Pal,
			Eliazar, and Reuveni}}]{PhysRevLett.122.020602}
	\bibinfo{author}{\bibfnamefont{A.}~\bibnamefont{Pal}},
	\bibinfo{author}{\bibfnamefont{I.}~\bibnamefont{Eliazar}}, \bibnamefont{and}
	\bibinfo{author}{\bibfnamefont{S.}~\bibnamefont{Reuveni}},
	\bibinfo{journal}{Phys. Rev. Lett.} \textbf{\bibinfo{volume}{122}},
	\bibinfo{pages}{020602} (\bibinfo{year}{2019}{\natexlab{a}}).
	
	\bibitem[{\citenamefont{Evans and Majumdar}(2014)}]{evans2014diffusion}
	\bibinfo{author}{\bibfnamefont{M.~R.} \bibnamefont{Evans}} \bibnamefont{and}
	\bibinfo{author}{\bibfnamefont{S.~N.} \bibnamefont{Majumdar}},
	\bibinfo{journal}{J. Phys. A: Math. Theor.} \textbf{\bibinfo{volume}{47}},
	\bibinfo{pages}{285001} (\bibinfo{year}{2014}).
	
	\bibitem[{\citenamefont{Christou and Schadschneider}(2015)}]{Christou2015}
	\bibinfo{author}{\bibfnamefont{C.}~\bibnamefont{Christou}} \bibnamefont{and}
	\bibinfo{author}{\bibfnamefont{A.}~\bibnamefont{Schadschneider}},
	\bibinfo{journal}{J. Phys. A: Math. Theor.} \textbf{\bibinfo{volume}{48}},
	\bibinfo{pages}{285003} (\bibinfo{year}{2015}).
	
	\bibitem[{\citenamefont{Durang et~al.}(2019)\citenamefont{Durang, Lee, Lizana,
			and Jeon}}]{Durang2019}
	\bibinfo{author}{\bibfnamefont{X.}~\bibnamefont{Durang}},
	\bibinfo{author}{\bibfnamefont{S.}~\bibnamefont{Lee}},
	\bibinfo{author}{\bibfnamefont{L.}~\bibnamefont{Lizana}}, \bibnamefont{and}
	\bibinfo{author}{\bibfnamefont{J.-H.} \bibnamefont{Jeon}},
	\bibinfo{journal}{J. Phys. A: Math. Theor.} \textbf{\bibinfo{volume}{52}},
	\bibinfo{pages}{224001} (\bibinfo{year}{2019}).
	
	\bibitem[{\citenamefont{Chen and Huang}(2022)}]{PhysRevE.105.034109}
	\bibinfo{author}{\bibfnamefont{H.}~\bibnamefont{Chen}} \bibnamefont{and}
	\bibinfo{author}{\bibfnamefont{F.}~\bibnamefont{Huang}},
	\bibinfo{journal}{Phys. Rev. E} \textbf{\bibinfo{volume}{105}},
	\bibinfo{pages}{034109} (\bibinfo{year}{2022}).
	
	\bibitem[{\citenamefont{Najeeb et~al.}(2026)\citenamefont{Najeeb, Pal, and
			Prasad}}]{3bc4-nlkt}
	\bibinfo{author}{\bibfnamefont{F.}~\bibnamefont{Najeeb}},
	\bibinfo{author}{\bibfnamefont{A.}~\bibnamefont{Pal}}, \bibnamefont{and}
	\bibinfo{author}{\bibfnamefont{V.~V.} \bibnamefont{Prasad}},
	\bibinfo{journal}{Phys. Rev. E} \textbf{\bibinfo{volume}{113}},
	\bibinfo{pages}{024104} (\bibinfo{year}{2026}).
	
	\bibitem[{\citenamefont{Boyer and Solis-Salas}(2014)}]{PhysRevLett.112.240601}
	\bibinfo{author}{\bibfnamefont{D.}~\bibnamefont{Boyer}} \bibnamefont{and}
	\bibinfo{author}{\bibfnamefont{C.}~\bibnamefont{Solis-Salas}},
	\bibinfo{journal}{Phys. Rev. Lett.} \textbf{\bibinfo{volume}{112}},
	\bibinfo{pages}{240601} (\bibinfo{year}{2014}).
	
	\bibitem[{\citenamefont{Kusmierz et~al.}(2014)\citenamefont{Kusmierz, Majumdar,
			Sabhapandit, and Schehr}}]{PhysRevLett.113.220602}
	\bibinfo{author}{\bibfnamefont{L.}~\bibnamefont{Kusmierz}},
	\bibinfo{author}{\bibfnamefont{S.~N.} \bibnamefont{Majumdar}},
	\bibinfo{author}{\bibfnamefont{S.}~\bibnamefont{Sabhapandit}},
	\bibnamefont{and} \bibinfo{author}{\bibfnamefont{G.}~\bibnamefont{Schehr}},
	\bibinfo{journal}{Phys. Rev. Lett.} \textbf{\bibinfo{volume}{113}},
	\bibinfo{pages}{220602} (\bibinfo{year}{2014}).
	
	\bibitem[{\citenamefont{Evans and Majumdar}(2018)}]{evans2018run}
	\bibinfo{author}{\bibfnamefont{M.~R.} \bibnamefont{Evans}} \bibnamefont{and}
	\bibinfo{author}{\bibfnamefont{S.~N.} \bibnamefont{Majumdar}},
	\bibinfo{journal}{J. Phys. A: Math. Theor.} \textbf{\bibinfo{volume}{51}},
	\bibinfo{pages}{475003} (\bibinfo{year}{2018}).
	
	\bibitem[{\citenamefont{Bressloff}(2020{\natexlab{a}})}]{bressloff2020occupation}
	\bibinfo{author}{\bibfnamefont{P.~C.} \bibnamefont{Bressloff}},
	\bibinfo{journal}{Phys. Rev. E} \textbf{\bibinfo{volume}{102}},
	\bibinfo{pages}{042135} (\bibinfo{year}{2020}{\natexlab{a}}).
	
	\bibitem[{\citenamefont{Kumar et~al.}(2020)\citenamefont{Kumar, Sadekar, and
			Basu}}]{kumar2020active}
	\bibinfo{author}{\bibfnamefont{V.}~\bibnamefont{Kumar}},
	\bibinfo{author}{\bibfnamefont{O.}~\bibnamefont{Sadekar}}, \bibnamefont{and}
	\bibinfo{author}{\bibfnamefont{U.}~\bibnamefont{Basu}},
	\bibinfo{journal}{Phys. Rev. E} \textbf{\bibinfo{volume}{102}},
	\bibinfo{pages}{052129} (\bibinfo{year}{2020}).
	
	\bibitem[{\citenamefont{Olsen et~al.}(2026)\citenamefont{Olsen, L{\"o}wen, and
			Caprini}}]{sytg-m4ms}
	\bibinfo{author}{\bibfnamefont{K.~S.} \bibnamefont{Olsen}},
	\bibinfo{author}{\bibfnamefont{H.}~\bibnamefont{L{\"o}wen}},
	\bibnamefont{and} \bibinfo{author}{\bibfnamefont{L.}~\bibnamefont{Caprini}},
	\bibinfo{journal}{Phys. Rev. Lett.} \textbf{\bibinfo{volume}{137}},
	\bibinfo{pages}{057102} (\bibinfo{year}{2026}).
	
	\bibitem[{\citenamefont{Wang et~al.}(2021)\citenamefont{Wang, Cherstvy, Kantz,
			Metzler, and Sokolov}}]{PhysRevE.104.024105}
	\bibinfo{author}{\bibfnamefont{W.}~\bibnamefont{Wang}},
	\bibinfo{author}{\bibfnamefont{A.~G.} \bibnamefont{Cherstvy}},
	\bibinfo{author}{\bibfnamefont{H.}~\bibnamefont{Kantz}},
	\bibinfo{author}{\bibfnamefont{R.}~\bibnamefont{Metzler}}, \bibnamefont{and}
	\bibinfo{author}{\bibfnamefont{I.~M.} \bibnamefont{Sokolov}},
	\bibinfo{journal}{Phys. Rev. E} \textbf{\bibinfo{volume}{104}},
	\bibinfo{pages}{024105} (\bibinfo{year}{2021}).
	
	\bibitem[{\citenamefont{Vinod et~al.}(2022)\citenamefont{Vinod, Cherstvy, Wang,
			Metzler, and Sokolov}}]{PhysRevE.105.L012106}
	\bibinfo{author}{\bibfnamefont{D.}~\bibnamefont{Vinod}},
	\bibinfo{author}{\bibfnamefont{A.~G.} \bibnamefont{Cherstvy}},
	\bibinfo{author}{\bibfnamefont{W.}~\bibnamefont{Wang}},
	\bibinfo{author}{\bibfnamefont{R.}~\bibnamefont{Metzler}}, \bibnamefont{and}
	\bibinfo{author}{\bibfnamefont{I.~M.} \bibnamefont{Sokolov}},
	\bibinfo{journal}{Phys. Rev. E} \textbf{\bibinfo{volume}{105}},
	\bibinfo{pages}{L012106} (\bibinfo{year}{2022}).
	
	\bibitem[{\citenamefont{Singh}(2020)}]{singh2020random}
	\bibinfo{author}{\bibfnamefont{P.}~\bibnamefont{Singh}}, \bibinfo{journal}{J.
		Phys. A: Math. Theor.} \textbf{\bibinfo{volume}{53}}, \bibinfo{pages}{405005}
	(\bibinfo{year}{2020}).
	
	\bibitem[{\citenamefont{De~Bruyne et~al.}(2022)\citenamefont{De~Bruyne,
			Majumdar, and Schehr}}]{PhysRevLett.128.200603}
	\bibinfo{author}{\bibfnamefont{B.}~\bibnamefont{De~Bruyne}},
	\bibinfo{author}{\bibfnamefont{S.~N.} \bibnamefont{Majumdar}},
	\bibnamefont{and} \bibinfo{author}{\bibfnamefont{G.}~\bibnamefont{Schehr}},
	\bibinfo{journal}{Phys. Rev. Lett.} \textbf{\bibinfo{volume}{128}},
	\bibinfo{pages}{200603} (\bibinfo{year}{2022}).
	
	\bibitem[{\citenamefont{Gupta et~al.}(2014)\citenamefont{Gupta, Majumdar, and
			Schehr}}]{gupta2014fluctuating}
	\bibinfo{author}{\bibfnamefont{S.}~\bibnamefont{Gupta}},
	\bibinfo{author}{\bibfnamefont{S.~N.} \bibnamefont{Majumdar}},
	\bibnamefont{and} \bibinfo{author}{\bibfnamefont{G.}~\bibnamefont{Schehr}},
	\bibinfo{journal}{Phys. Rev. Lett.} \textbf{\bibinfo{volume}{112}},
	\bibinfo{pages}{220601} (\bibinfo{year}{2014}).
	
	\bibitem[{\citenamefont{Meylahn et~al.}(2015)\citenamefont{Meylahn,
			Sabhapandit, and Touchette}}]{meylahn2015large}
	\bibinfo{author}{\bibfnamefont{J.~M.} \bibnamefont{Meylahn}},
	\bibinfo{author}{\bibfnamefont{S.}~\bibnamefont{Sabhapandit}},
	\bibnamefont{and}
	\bibinfo{author}{\bibfnamefont{H.}~\bibnamefont{Touchette}},
	\bibinfo{journal}{Phys. Rev. E} \textbf{\bibinfo{volume}{92}},
	\bibinfo{pages}{062148} (\bibinfo{year}{2015}).
	
	\bibitem[{\citenamefont{Kumar and Pal}(2023)}]{PhysRevLett.130.157101}
	\bibinfo{author}{\bibfnamefont{A.}~\bibnamefont{Kumar}} \bibnamefont{and}
	\bibinfo{author}{\bibfnamefont{A.}~\bibnamefont{Pal}},
	\bibinfo{journal}{Phys. Rev. Lett.} \textbf{\bibinfo{volume}{130}},
	\bibinfo{pages}{157101} (\bibinfo{year}{2023}).
	
	\bibitem[{\citenamefont{Singh and Pal}(2021)}]{PhysRevE.103.052119}
	\bibinfo{author}{\bibfnamefont{P.}~\bibnamefont{Singh}} \bibnamefont{and}
	\bibinfo{author}{\bibfnamefont{A.}~\bibnamefont{Pal}},
	\bibinfo{journal}{Phys. Rev. E} \textbf{\bibinfo{volume}{103}},
	\bibinfo{pages}{052119} (\bibinfo{year}{2021}).
	
	\bibitem[{\citenamefont{Guo et~al.}(2023)\citenamefont{Guo, Yan, and
			Chen}}]{guo2023extremal}
	\bibinfo{author}{\bibfnamefont{W.}~\bibnamefont{Guo}},
	\bibinfo{author}{\bibfnamefont{H.}~\bibnamefont{Yan}}, \bibnamefont{and}
	\bibinfo{author}{\bibfnamefont{H.}~\bibnamefont{Chen}},
	\bibinfo{journal}{Phys. Rev. E} \textbf{\bibinfo{volume}{108}},
	\bibinfo{pages}{044115} (\bibinfo{year}{2023}).
	
	\bibitem[{\citenamefont{Blumer et~al.}(2022)\citenamefont{Blumer, Reuveni, and
			Hirshberg}}]{blumer2022stochastic}
	\bibinfo{author}{\bibfnamefont{O.}~\bibnamefont{Blumer}},
	\bibinfo{author}{\bibfnamefont{S.}~\bibnamefont{Reuveni}}, \bibnamefont{and}
	\bibinfo{author}{\bibfnamefont{B.}~\bibnamefont{Hirshberg}},
	\bibinfo{journal}{J. Phys. Chem. Lett.} \textbf{\bibinfo{volume}{13}},
	\bibinfo{pages}{11230} (\bibinfo{year}{2022}).
	
	\bibitem[{\citenamefont{Blumer et~al.}(2024)\citenamefont{Blumer, Reuveni, and
			Hirshberg}}]{blumer2024combining}
	\bibinfo{author}{\bibfnamefont{O.}~\bibnamefont{Blumer}},
	\bibinfo{author}{\bibfnamefont{S.}~\bibnamefont{Reuveni}}, \bibnamefont{and}
	\bibinfo{author}{\bibfnamefont{B.}~\bibnamefont{Hirshberg}},
	\bibinfo{journal}{Nat. Commun.} \textbf{\bibinfo{volume}{15}},
	\bibinfo{pages}{240} (\bibinfo{year}{2024}).
	
	\bibitem[{\citenamefont{Tal-Friedman et~al.}(2020)\citenamefont{Tal-Friedman,
			Pal, Sekhon, Reuveni, and Roichman}}]{tal2020experimental}
	\bibinfo{author}{\bibfnamefont{O.}~\bibnamefont{Tal-Friedman}},
	\bibinfo{author}{\bibfnamefont{A.}~\bibnamefont{Pal}},
	\bibinfo{author}{\bibfnamefont{A.}~\bibnamefont{Sekhon}},
	\bibinfo{author}{\bibfnamefont{S.}~\bibnamefont{Reuveni}}, \bibnamefont{and}
	\bibinfo{author}{\bibfnamefont{Y.}~\bibnamefont{Roichman}},
	\bibinfo{journal}{J. Phys. Chem. Lett.} \textbf{\bibinfo{volume}{11}},
	\bibinfo{pages}{7350} (\bibinfo{year}{2020}).
	
	\bibitem[{\citenamefont{Besga et~al.}(2020)\citenamefont{Besga, Bovon,
			Petrosyan, Majumdar, and Ciliberto}}]{besga2020optimal}
	\bibinfo{author}{\bibfnamefont{B.}~\bibnamefont{Besga}},
	\bibinfo{author}{\bibfnamefont{A.}~\bibnamefont{Bovon}},
	\bibinfo{author}{\bibfnamefont{A.}~\bibnamefont{Petrosyan}},
	\bibinfo{author}{\bibfnamefont{S.~N.} \bibnamefont{Majumdar}},
	\bibnamefont{and}
	\bibinfo{author}{\bibfnamefont{S.}~\bibnamefont{Ciliberto}},
	\bibinfo{journal}{Phys. Rev. Res.} \textbf{\bibinfo{volume}{2}},
	\bibinfo{pages}{032029} (\bibinfo{year}{2020}).
	
	\bibitem[{\citenamefont{Besga et~al.}(2021)\citenamefont{Besga, Faisant,
			Petrosyan, Ciliberto, and Majumdar}}]{besga2021dynamical}
	\bibinfo{author}{\bibfnamefont{B.}~\bibnamefont{Besga}},
	\bibinfo{author}{\bibfnamefont{F.}~\bibnamefont{Faisant}},
	\bibinfo{author}{\bibfnamefont{A.}~\bibnamefont{Petrosyan}},
	\bibinfo{author}{\bibfnamefont{S.}~\bibnamefont{Ciliberto}},
	\bibnamefont{and} \bibinfo{author}{\bibfnamefont{S.~N.}
		\bibnamefont{Majumdar}}, \bibinfo{journal}{Phys. Rev. E}
	\textbf{\bibinfo{volume}{104}}, \bibinfo{pages}{L012102}
	(\bibinfo{year}{2021}).
	
	\bibitem[{\citenamefont{Goerlich et~al.}(2025)\citenamefont{Goerlich, Li,
			Pires, Hervieux, Manfredi, and Genet}}]{goerlich2025taming}
	\bibinfo{author}{\bibfnamefont{R.}~\bibnamefont{Goerlich}},
	\bibinfo{author}{\bibfnamefont{M.}~\bibnamefont{Li}},
	\bibinfo{author}{\bibfnamefont{L.~B.} \bibnamefont{Pires}},
	\bibinfo{author}{\bibfnamefont{P.-A.} \bibnamefont{Hervieux}},
	\bibinfo{author}{\bibfnamefont{G.}~\bibnamefont{Manfredi}}, \bibnamefont{and}
	\bibinfo{author}{\bibfnamefont{C.}~\bibnamefont{Genet}},
	\bibinfo{journal}{Phys. Rev. E} \textbf{\bibinfo{volume}{112}},
	\bibinfo{pages}{064116} (\bibinfo{year}{2025}).
	
	\bibitem[{\citenamefont{Evans and Majumdar}(2019)}]{evans2019effects}
	\bibinfo{author}{\bibfnamefont{M.~R.} \bibnamefont{Evans}} \bibnamefont{and}
	\bibinfo{author}{\bibfnamefont{S.~N.} \bibnamefont{Majumdar}},
	\bibinfo{journal}{J. Phys. A: Math. Theor.} \textbf{\bibinfo{volume}{52}},
	\bibinfo{pages}{01LT01} (\bibinfo{year}{2019}).
	
	\bibitem[{\citenamefont{Pal et~al.}(2019{\natexlab{b}})\citenamefont{Pal,
			Ku{\'s}mierz, and Reuveni}}]{pal2019invariants}
	\bibinfo{author}{\bibfnamefont{A.}~\bibnamefont{Pal}},
	\bibinfo{author}{\bibfnamefont{{\L}.}~\bibnamefont{Ku{\'s}mierz}},
	\bibnamefont{and} \bibinfo{author}{\bibfnamefont{S.}~\bibnamefont{Reuveni}},
	\bibinfo{journal}{New J. Phys.} \textbf{\bibinfo{volume}{21}},
	\bibinfo{pages}{113024} (\bibinfo{year}{2019}{\natexlab{b}}).
	
	\bibitem[{\citenamefont{Pal et~al.}(2019{\natexlab{c}})\citenamefont{Pal,
			Ku{\'s}mierz, and Reuveni}}]{pal2019time}
	\bibinfo{author}{\bibfnamefont{A.}~\bibnamefont{Pal}},
	\bibinfo{author}{\bibfnamefont{{\L}.}~\bibnamefont{Ku{\'s}mierz}},
	\bibnamefont{and} \bibinfo{author}{\bibfnamefont{S.}~\bibnamefont{Reuveni}},
	\bibinfo{journal}{Phys. Rev. E} \textbf{\bibinfo{volume}{100}},
	\bibinfo{pages}{040101(R)} (\bibinfo{year}{2019}{\natexlab{c}}).
	
	\bibitem[{\citenamefont{Maso-Puigdollosas
			et~al.}(2019)\citenamefont{Maso-Puigdollosas, Campos, and
			Mendez}}]{maso2019transport}
	\bibinfo{author}{\bibfnamefont{A.}~\bibnamefont{Maso-Puigdollosas}},
	\bibinfo{author}{\bibfnamefont{D.}~\bibnamefont{Campos}}, \bibnamefont{and}
	\bibinfo{author}{\bibfnamefont{V.}~\bibnamefont{Mendez}},
	\bibinfo{journal}{Phys. Rev. E} \textbf{\bibinfo{volume}{100}},
	\bibinfo{pages}{042104} (\bibinfo{year}{2019}).
	
	\bibitem[{\citenamefont{Bodrova and
			Sokolov}(2020{\natexlab{a}})}]{bodrova2020resetting}
	\bibinfo{author}{\bibfnamefont{A.~S.} \bibnamefont{Bodrova}} \bibnamefont{and}
	\bibinfo{author}{\bibfnamefont{I.~M.} \bibnamefont{Sokolov}},
	\bibinfo{journal}{Phys. Rev. E} \textbf{\bibinfo{volume}{101}},
	\bibinfo{pages}{052130} (\bibinfo{year}{2020}{\natexlab{a}}).
	
	\bibitem[{\citenamefont{Bodrova and
			Sokolov}(2020{\natexlab{b}})}]{bodrova2020brownian}
	\bibinfo{author}{\bibfnamefont{A.~S.} \bibnamefont{Bodrova}} \bibnamefont{and}
	\bibinfo{author}{\bibfnamefont{I.~M.} \bibnamefont{Sokolov}},
	\bibinfo{journal}{Phys. Rev. E} \textbf{\bibinfo{volume}{102}},
	\bibinfo{pages}{032129} (\bibinfo{year}{2020}{\natexlab{b}}).
	
	\bibitem[{\citenamefont{Pal et~al.}(2020)\citenamefont{Pal, Ku{\'s}mierz, and
			Reuveni}}]{pal2020search}
	\bibinfo{author}{\bibfnamefont{A.}~\bibnamefont{Pal}},
	\bibinfo{author}{\bibfnamefont{{\L}.}~\bibnamefont{Ku{\'s}mierz}},
	\bibnamefont{and} \bibinfo{author}{\bibfnamefont{S.}~\bibnamefont{Reuveni}},
	\bibinfo{journal}{Phys. Rev. Res.} \textbf{\bibinfo{volume}{2}},
	\bibinfo{pages}{043174} (\bibinfo{year}{2020}).
	
	\bibitem[{\citenamefont{Gupta et~al.}(2020{\natexlab{a}})\citenamefont{Gupta,
			Plata, Kundu, and Pal}}]{gupta2020stochastic}
	\bibinfo{author}{\bibfnamefont{D.}~\bibnamefont{Gupta}},
	\bibinfo{author}{\bibfnamefont{C.~A.} \bibnamefont{Plata}},
	\bibinfo{author}{\bibfnamefont{A.}~\bibnamefont{Kundu}}, \bibnamefont{and}
	\bibinfo{author}{\bibfnamefont{A.}~\bibnamefont{Pal}}, \bibinfo{journal}{J.
		Phys. A: Math. Theor.} \textbf{\bibinfo{volume}{54}}, \bibinfo{pages}{025003}
	(\bibinfo{year}{2020}{\natexlab{a}}).
	
	\bibitem[{\citenamefont{Gupta et~al.}(2021)\citenamefont{Gupta, Plata, Pal, and
			Kundu}}]{gupta2021resetting}
	\bibinfo{author}{\bibfnamefont{D.}~\bibnamefont{Gupta}},
	\bibinfo{author}{\bibfnamefont{C.~A.} \bibnamefont{Plata}},
	\bibinfo{author}{\bibfnamefont{A.}~\bibnamefont{Pal}}, \bibnamefont{and}
	\bibinfo{author}{\bibfnamefont{A.}~\bibnamefont{Kundu}}, \bibinfo{journal}{J.
		Stat. Mech.: Theory Exp.} \textbf{\bibinfo{volume}{2021}},
	\bibinfo{pages}{043202} (\bibinfo{year}{2021}).
	
	\bibitem[{\citenamefont{Biswas et~al.}(2024)\citenamefont{Biswas, Kundu, and
			Pal}}]{PhysRevE.110.L042101}
	\bibinfo{author}{\bibfnamefont{A.}~\bibnamefont{Biswas}},
	\bibinfo{author}{\bibfnamefont{A.}~\bibnamefont{Kundu}}, \bibnamefont{and}
	\bibinfo{author}{\bibfnamefont{A.}~\bibnamefont{Pal}},
	\bibinfo{journal}{Phys. Rev. E} \textbf{\bibinfo{volume}{110}},
	\bibinfo{pages}{L042101} (\bibinfo{year}{2024}).
	
	\bibitem[{\citenamefont{Fuchs et~al.}(2016)\citenamefont{Fuchs, Goldt, and
			Seifert}}]{fuchs2016stochastic}
	\bibinfo{author}{\bibfnamefont{J.}~\bibnamefont{Fuchs}},
	\bibinfo{author}{\bibfnamefont{S.}~\bibnamefont{Goldt}}, \bibnamefont{and}
	\bibinfo{author}{\bibfnamefont{U.}~\bibnamefont{Seifert}},
	\bibinfo{journal}{EPL (Europhysics Letters)} \textbf{\bibinfo{volume}{113}},
	\bibinfo{pages}{60009} (\bibinfo{year}{2016}).
	
	\bibitem[{\citenamefont{Sunil et~al.}(2023)\citenamefont{Sunil, Blythe, Evans,
			and Majumdar}}]{sunil2023cost}
	\bibinfo{author}{\bibfnamefont{J.~C.} \bibnamefont{Sunil}},
	\bibinfo{author}{\bibfnamefont{R.~A.} \bibnamefont{Blythe}},
	\bibinfo{author}{\bibfnamefont{M.~R.} \bibnamefont{Evans}}, \bibnamefont{and}
	\bibinfo{author}{\bibfnamefont{S.~N.} \bibnamefont{Majumdar}},
	\bibinfo{journal}{J. Phys. A: Math. Theor.} \textbf{\bibinfo{volume}{56}},
	\bibinfo{pages}{395001} (\bibinfo{year}{2023}).
	
	\bibitem[{\citenamefont{Olsen et~al.}(2024)\citenamefont{Olsen, Gupta, Mori,
			and Krishnamurthy}}]{olsen2024thermodynamic}
	\bibinfo{author}{\bibfnamefont{K.~S.} \bibnamefont{Olsen}},
	\bibinfo{author}{\bibfnamefont{D.}~\bibnamefont{Gupta}},
	\bibinfo{author}{\bibfnamefont{F.}~\bibnamefont{Mori}}, \bibnamefont{and}
	\bibinfo{author}{\bibfnamefont{S.}~\bibnamefont{Krishnamurthy}},
	\bibinfo{journal}{Phys. Rev. Res.} \textbf{\bibinfo{volume}{6}},
	\bibinfo{pages}{033343} (\bibinfo{year}{2024}).
	
	\bibitem[{\citenamefont{Goerlich et~al.}(2026)\citenamefont{Goerlich, Olsen,
			L{\"o}wen, and Roichman}}]{goerlich2026time}
	\bibinfo{author}{\bibfnamefont{R.}~\bibnamefont{Goerlich}},
	\bibinfo{author}{\bibfnamefont{K.~S.} \bibnamefont{Olsen}},
	\bibinfo{author}{\bibfnamefont{H.}~\bibnamefont{L{\"o}wen}},
	\bibnamefont{and} \bibinfo{author}{\bibfnamefont{Y.}~\bibnamefont{Roichman}},
	\bibinfo{journal}{Phys. Rev. E} \textbf{\bibinfo{volume}{113}},
	\bibinfo{pages}{014103} (\bibinfo{year}{2026}).
	
	\bibitem[{\citenamefont{Seifert}(2012)}]{seifert2012stochastic}
	\bibinfo{author}{\bibfnamefont{U.}~\bibnamefont{Seifert}},
	\bibinfo{journal}{Reports on Progress in Physics}
	\textbf{\bibinfo{volume}{75}}, \bibinfo{pages}{126001}
	(\bibinfo{year}{2012}).
	
	\bibitem[{\citenamefont{Gupta et~al.}(2020{\natexlab{b}})\citenamefont{Gupta,
			Plata, and Pal}}]{gupta2020work}
	\bibinfo{author}{\bibfnamefont{D.}~\bibnamefont{Gupta}},
	\bibinfo{author}{\bibfnamefont{C.~A.} \bibnamefont{Plata}}, \bibnamefont{and}
	\bibinfo{author}{\bibfnamefont{A.}~\bibnamefont{Pal}},
	\bibinfo{journal}{Phys. Rev. Lett.} \textbf{\bibinfo{volume}{124}},
	\bibinfo{pages}{110608} (\bibinfo{year}{2020}{\natexlab{b}}).
	
	\bibitem[{\citenamefont{Gupta and Plata}(2022)}]{gupta2022work}
	\bibinfo{author}{\bibfnamefont{D.}~\bibnamefont{Gupta}} \bibnamefont{and}
	\bibinfo{author}{\bibfnamefont{C.~A.} \bibnamefont{Plata}},
	\bibinfo{journal}{New J. Phys.} \textbf{\bibinfo{volume}{24}},
	\bibinfo{pages}{113034} (\bibinfo{year}{2022}).
	
	\bibitem[{\citenamefont{Pal et~al.}(2023)\citenamefont{Pal, Pal, Park, and
			Lee}}]{pal2023thermodynamic}
	\bibinfo{author}{\bibfnamefont{P.~S.} \bibnamefont{Pal}},
	\bibinfo{author}{\bibfnamefont{A.}~\bibnamefont{Pal}},
	\bibinfo{author}{\bibfnamefont{H.}~\bibnamefont{Park}}, \bibnamefont{and}
	\bibinfo{author}{\bibfnamefont{J.~S.} \bibnamefont{Lee}},
	\bibinfo{journal}{Phys. Rev. E} \textbf{\bibinfo{volume}{108}},
	\bibinfo{pages}{044117} (\bibinfo{year}{2023}).
	
	\bibitem[{\citenamefont{Singh}(2024)}]{singh2024emerging}
	\bibinfo{author}{\bibfnamefont{P.}~\bibnamefont{Singh}}, \bibinfo{journal}{New
		J. Phys.} \textbf{\bibinfo{volume}{26}}, \bibinfo{pages}{103014}
	(\bibinfo{year}{2024}).
	
	\bibitem[{\citenamefont{Bressloff}(2020{\natexlab{b}})}]{PhysRevE.102.022115}
	\bibinfo{author}{\bibfnamefont{P.~C.} \bibnamefont{Bressloff}},
	\bibinfo{journal}{Phys. Rev. E} \textbf{\bibinfo{volume}{102}},
	\bibinfo{pages}{022115} (\bibinfo{year}{2020}{\natexlab{b}}).
	
	\bibitem[{\citenamefont{Pal and
			Prasad}(2019{\natexlab{a}})}]{PhysRevE.99.032123}
	\bibinfo{author}{\bibfnamefont{A.}~\bibnamefont{Pal}} \bibnamefont{and}
	\bibinfo{author}{\bibfnamefont{V.~V.} \bibnamefont{Prasad}},
	\bibinfo{journal}{Phys. Rev. E} \textbf{\bibinfo{volume}{99}},
	\bibinfo{pages}{032123} (\bibinfo{year}{2019}{\natexlab{a}}).
	
	\bibitem[{\citenamefont{Reuveni}(2016)}]{PhysRevLett.116.170601}
	\bibinfo{author}{\bibfnamefont{S.}~\bibnamefont{Reuveni}},
	\bibinfo{journal}{Phys. Rev. Lett.} \textbf{\bibinfo{volume}{116}},
	\bibinfo{pages}{170601} (\bibinfo{year}{2016}).
	
	\bibitem[{\citenamefont{Chatterjee et~al.}(2018)\citenamefont{Chatterjee,
			Christou, and Schadschneider}}]{PhysRevE.97.062106}
	\bibinfo{author}{\bibfnamefont{A.}~\bibnamefont{Chatterjee}},
	\bibinfo{author}{\bibfnamefont{C.}~\bibnamefont{Christou}}, \bibnamefont{and}
	\bibinfo{author}{\bibfnamefont{A.}~\bibnamefont{Schadschneider}},
	\bibinfo{journal}{Phys. Rev. E} \textbf{\bibinfo{volume}{97}},
	\bibinfo{pages}{062106} (\bibinfo{year}{2018}).
	
	\bibitem[{\citenamefont{Pal and
			Prasad}(2019{\natexlab{b}})}]{PhysRevResearch.1.032001}
	\bibinfo{author}{\bibfnamefont{A.}~\bibnamefont{Pal}} \bibnamefont{and}
	\bibinfo{author}{\bibfnamefont{V.~V.} \bibnamefont{Prasad}},
	\bibinfo{journal}{Phys. Rev. Research} \textbf{\bibinfo{volume}{1}},
	\bibinfo{pages}{032001} (\bibinfo{year}{2019}{\natexlab{b}}).
	
	\bibitem[{\citenamefont{Majumdar}(2005)}]{majumdar2005brownian}
	\bibinfo{author}{\bibfnamefont{S.~N.} \bibnamefont{Majumdar}},
	\bibinfo{journal}{Current Science} \textbf{\bibinfo{volume}{89}},
	\bibinfo{pages}{2076} (\bibinfo{year}{2005}).
	
	\bibitem[{\citenamefont{Singh and Pal}(2022)}]{singh2022first}
	\bibinfo{author}{\bibfnamefont{P.}~\bibnamefont{Singh}} \bibnamefont{and}
	\bibinfo{author}{\bibfnamefont{A.}~\bibnamefont{Pal}}, \bibinfo{journal}{J.
		Phys. A: Math. Theor.} \textbf{\bibinfo{volume}{55}}, \bibinfo{pages}{234001}
	(\bibinfo{year}{2022}).
	
	\bibitem[{\citenamefont{Sekimoto}(2010)}]{sekimoto2010stochastic}
	\bibinfo{author}{\bibfnamefont{K.}~\bibnamefont{Sekimoto}},
	\emph{\bibinfo{title}{Stochastic Energetics}} (\bibinfo{publisher}{Springer},
	\bibinfo{address}{Berlin}, \bibinfo{year}{2010}).
	
\end{thebibliography}

\end{document}